\documentclass[useAMS,usenatbib]{mn2e}

\usepackage{lscape}
\usepackage{graphicx}
\usepackage{natbib}
\usepackage{graphics}

\title[Galaxy morphology in WINGS]{Morphology of galaxies in the WINGS clusters}
\author[G. Fasano et al.]{
\parbox[t]{\textwidth}{
G. Fasano$^{1}$\thanks{E-mail:giovanni.fasano@oapd.inaf.it}, 
E. Vanzella$^{2}$,
A. Dressler$^{3}$, 
B.M. Poggianti$^{1}$, 
M. Moles$^{4}$, 
D. Bettoni$^{1}$, 
T. Valentinuzzi$^{5}$, 
A. Moretti$^{1}$,
M. D'Onofrio$^{5}$, 
J. Varela,
W.J. Couch$^{6}$, 
P. Kj\ae rgaard$^{7}$,
J. Fritz$^{8}$,
A. Omizzolo$^{1,9}$ and 
A. Cava$^{10}$.}\\
\\
$^{1}$INAF, Osservatorio Astronomico di Padova, Vicolo Osservatorio 5, 35122 Padova, Italy\\
$^{2}$INAF, Osservatorio Astronomico di Trieste, Via Tiepolo 11, 34143 Trieste, Italy\\
$^{3}$The Observatories of the Carnegie Institution of Washington, Pasadena, USA\\
$^{4}$Instituto de Astrofisica de Andalucia, Camino Bajo de Huetor 50, 18008 Granada, Spain\\
$^{5}$Dip. Astronomia, Universit\`a di Padova, Vicolo dell'Osservatorio 2, 35122 Padova, Italy\\
$^{6}$Center for Astrophysics, Swinburne University of Technology, Australia\\
$^{7}$Copenhagen University Observatory, Niels Bohr Institute for Astronomy, Denmark\\
$^{8}$Sterrenkundig Observatorium, Universiteit Gent, Krijgslaan 281 S9, B-9000 Gent, Belgium\\
$^{9}$Specola Vaticana, 00120 Stato Citt\`a del Vaticano\\
$^{10}$Departamento de Astrof\'{\i}sica, Facultad de
CC. F\'{\i}isicas, Universidad Complutense de Madrid, E-28040, Madrid, Spain
}
\begin{document}

\date{Accepted .... Received .....; in original form .....}

\pagerange{\pageref{firstpage}--\pageref{lastpage}} \pubyear{2011}

\maketitle

\label{firstpage}

\begin{abstract}

  We present the morphological catalog of galaxies in nearby
  clusters of the WINGS survey \citep{fasa06}. The catalog contains a
  total number of 39923 galaxies, for which we provide the automatic
  estimates of the morphological type applying the purposely devised
  tool MORPHOT to the V-band WINGS imaging. For $\sim$3000 galaxies
  we also provide visual estimates of the morphological types. A
  substantial part of the paper is devoted to the description of the
  MORPHOT tool, whose application is limited, at least for the moment,
  to the WINGS imaging only. The approach of the tool to the
  automation of morphological classification is a non parametric and
  fully empirical one. In particular, MORPHOT exploits 21
  morphological diagnostics, directly and easily computable from the
  galaxy image, to provide two independent classifications: one based
  on a Maximum Likelihood (ML), semi-analytical technique, the other
  one on a Neural Network (NN) machine. A suitably selected sample of
  $\sim$1000 visually classified WINGS galaxies is used to calibrate
  the diagnostics for the ML estimator and as a training set in the NN
  machine. The final morphological estimator combines the two
  techniques and proves to be effective both when applied to an
  additional test sample of $\sim$1000 visually classified WINGS
  galaxies and when compared with small samples of SDSS galaxies
  visually classified by \citet{fuku07} and \citet{nair10}.  Finally,
  besides the galaxy morphology distribution (corrected for field
  contamination) in the WINGS clusters, we present the ellipticity
  ($\varepsilon$), color (B-V) and Sersic index ($n$) distributions
  for different morphological types, as well as the morphological
  fractions as a function of the clustercentric distance (in units of
  $R_{200}$).

\end{abstract}


\section{Introduction} \label{secintro}

The \textit{WIde-field Nearby Galaxy-clusters Survey}
\citep[WINGS,][]{fasa06} has gathered wide-field,
photometric data, in the optical bands B and V \citep{vare09}, of
several hundred thousand galaxies in the fields of 76 nearby clusters
(0.04$\le$z$\le$0.07), selected from three X-ray flux limited samples
compiled from ROSAT All-Sky Survey data \citep{ebel96,ebel98,ebel00}.
The observations in the optical bands have been obtained with the
WFC@INT and with the WFI@MPG/ESO-2.2 cameras for the northern and
southern clusters, respectively.  Follow-ups of the optical WINGS
survey include medium-resolution, multi-fiber spectra of $\sim$6000
galaxies in 48 WINGS clusters \citep{cava09}, wide-field imaging in
the NIR bands J and K of $\sim$10$^6$ galaxies in 28 WINGS clusters
\citep{vale09} and U-band, wide-field imaging of 18 WINGS clusters
\citep{omiz11}. Lastly, a narrow H$_\alpha$-band photometric survey is
presently ongoing on a sub-set of the WINGS cluster sample.

The WINGS survey was conceived in 2000, mainly with the aim of making
up for the odd situation for which, while a large amount of high
quality morphological data for distant clusters were already available
from HST imaging \citep{couc94,pasc95,oeml97,kels97,couc98,lubi98},
high quality CCD data were almost lacking for large samples of nearby
clusters. Actually, the very selection of the WINGS cluster sample,
as well as the choice of the telescopes and the observational
constraints of the optical survey, were performed in order to meet the
requirements needed by the main original task (morphology of galaxies
in clusters), in terms of absolute field of view ($>$1.6~Mpc) and
spatial resolution (1$^{\prime\prime}<$1.3Kpc).

A recent, comprehensive review of the various aspects and issues
linked to galaxy morphology can be found in \citet{buta11}.  Till a
dozen years ago the morphological type estimate of galaxies has been
obtained just by visual inspection of photografic plates or CCD
frames. A few attempts were actually made in the 90's to obtain
automated morphological classification of galaxies \citep[with neural
networks and self-organizing maps;][]{naim95,naim97}, but they
remained isolated.  In the last decade, the sudden availability of
CCD mosaics has made it no longer feasible to conduct morphological
classifications by eye, since one has often to deal
with wide and/or deep fields, each one containing thousands galaxies.
This has triggered a number of papers proposing different tools for
automatic morphology estimate of large galaxy samples.

\begin{figure}
   \vspace*{-2truecm}
   \centering
   \includegraphics[width=8truecm,trim=150 0 150 0]{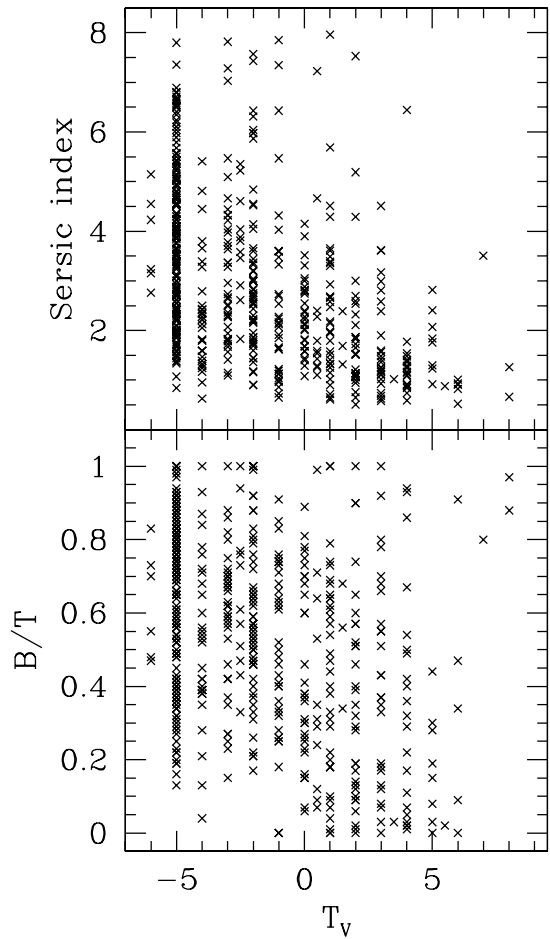}
   \vspace*{-4.5truecm}
      \caption{
        Visual morphology versus Sersic's index (top panel) and bulge
        fraction (bottom panel) for the 527 galaxies in common between
        the MORPHOT calibration sample (see Section~\ref{calsamp}) and
        the WINGS-GASP2D sample \citep{sanc11}.}
         \label{BDSer}
   \end{figure}

There are basically two alternative approaches to the problem of
automatic morphological classification of galaxies. The first one
exploits the parametrization of their radial light profiles \citep[see
for instance][]{guti04,truj04,sain05,tasc05,ornd05,ravi06,cass10}.  In
this case, the most commonly used morphological diagnostics are the
bulge fraction ({\it B/T}) and the Sersic's index ({\it n}). Several
tools have been devised to obtain, in (semi)automatic mode, the
best-fit parameters of the analytical laws used to represent the light
distribution of galaxies. Among the others, we mention GIM2D
\citep{sima98}, GALFIT \citep{peng02}, GASPHOT \citep{pign06}, 2DPHOT
\citep{laba08}, GASP2D \citep{mend08} and GALPATH \citep{yoon09}.
However, this approach to the morphological classification of galaxies
presents two serious drawbacks: (i) the analytical components derived
from the formal best-fitting of galaxy light profiles (usually
exponential and Sersic laws) often do not correspond to real physical
components \citep[disk and bulge; see for instance][]{tasc05}; (ii)
the correlations between these diagnostics ({\it B/T} and {\it n}) and
the visual morphological type are weak and show a high degree of
degeneracy, especially for early-type galaxies \citep[see 
Figure~\ref{BDSer}, see also][]{sanc04,pign06}. These
drawbacks reflect the fact that structure and morphology 
of galaxies are intrinsically different concepts
\citep[see][]{vand08}.  In fact, while the first one is a global
property that can be described by means of simple analytical laws and
leaves mostly aside the problems connected with image texture, $S/N$
ratio and resolution (apart from the convolution with the local PSFs),
the second one mainly deals with pixel-scale behaviours and features
which, in default of human eyes, make difficult any quantitative
description.

The alternative, non-parametric approach tries to face the problem
relying upon various diagnostics, directly computable from the digital
post-stamp images of galaxies, which are empirically found to
correlate with the visual morphological estimates. The non-parametric
tools can be in turn divided in two main categories: (i) those using a
few (two or three) diagnostics and the relative two- or
three-dimension space to try to segregate galaxies with different
morphological types; (ii) those using Neural Networks (or some other
sharp methodology) to combine many diagnostics, thus drawing 
a final, quantitative estimate of the morphological
type. Among the tools belonging to the first category, besides {\bf the pioneering 
diagnostic devised by \citet[][concentration vs. asymmetry]{abra96} and} the
popular CAS (Concentration/Asymmetry/clumpinesS) diagnostic set
\citep{cons03}, it is worth mentioning those proposed by {\bf \citet[][Gini Coefficient]{abra03} and}
\citet[][M20 coefficient]{lotz04}, \citet[][concentration and 
asymmetry at different wavelenghts]{laug05}, \citet[][concentration and 
coarseness coefficients]{yama05}, \citet{mena06}, \citet{vand08}
and \citet{pett09}. To the second category belong the tools
devised by \citet[][Fourier analysis of the images]{odew02}, \citet{ball04},
\citet{gode04}, \citet{dela04}, \citet[][shapelet analysis]{kell04}, 
\citet{moor06}, \citet[][ZEST]{scar07}, \citet{huer08} and \citet{sham09}.
A mixed approach ({\it B/T} decomposition $+$ non parametric diagnostics) has 
been tried by \citet{rahm04}, \citet{chen11} and \citet[][PyMorph]{vikr10}.

The non-parametric approach seems to be more effective than
the parametric one in estimating the morphological type of galaxies
\citep{hatz05} and has been claimed to be even able to
compare with visual estimates as far as the intrinsic scatter and the
robustness of the results are concerned \citep{odew02,bell04,huer08}. 
However, a common limitation of the non-parametric tools available in
the literature is the scarce ability of separating S0s from elliptical
galaxies, which is actually an important issue when dealing with
galaxy evolution in clusters
\citep{dres97,fasa00,treu03,post05,desa07,pogg09}.

In this paper we describe a new automatic, non-parametric tool for the
morphological type estimate of large galaxy samples
\citep[MORPHOT][]{fasa07}, which is in fact able to separate Es from
S0s in a majority of cases.  MORPHOT belongs to the second previously
mentioned category of non parametric tools. It starts with a set of 21
suitably devised morphological diagnostics, and combines them in two
different (independent) ways, thus producing the final
morphological type (and the relative confidence interval) for each
galaxy in a given input catalog. We fine-tune MORPHOT for extensive
application to the WINGS cluster sample and present the catalogs of
the survey, which contain morphological types of $\sim$40000
galaxies. We stress that, although the basic methodology is robust for
any set of digital images of similar spatial resolution and dynamic
range, at this stage the tool does not pretend to have a general
validity, regardless of the observing conditions (telescope, detector,
seeing) and the galaxy sample (redshift). However, we will show that
it produces reliable results for the particular purposes of the
WINGS survey.

In Section~\ref{secvis} we report about the intrinsic reliabity of the
visual morphological classifications. In Section~\ref{sectool} we
describe in some detail the structure of MORPHOT and the various steps
of the tool's flow-chart. In Section~\ref{secperf} we analyse the
performances of MORPHOT on the WINGS galaxy sample. In
Section~\ref{secwings} we apply MORPHOT to the WINGS cluster galaxies,
present the WINGS catalogs of morphological types and briefly discuss
the main statistical properties of galaxy morphology in the WINGS
clusters. 
Section~\ref{secsum} summarizes the results and outlines the future
employment of the MORPHOT classifications. 
Throughout the paper we adopt the following cosmology:
$H_0$=70~kms$^{-1}$Mpc$^{-1}$, $\Omega_M$=0.3 and
$\Omega_\Lambda$=0.7.

\section{How reliable is the visual classification~?}\label{secvis}

{\bf After the pioneering attempt by \citet{rayn20} to provide a morphological classification
of spiral nebulae and} since the first, definite understanding by Edwin~\citet{hubb25} of the
extragalactic nature of many nebulae, a number of different
classification schemes have been proposed for the galaxy
morphology. The original Hubble's sequence
\citep[][spirals/elongated/globular/irregulars]{hubb22} was improved
by the author himself, first introducing the concept of {\it tuning
  fork} to distinguish between normal and barred disk galaxies
\citep{hubb26}, then defining the S0 morphological type \citep{hubb36}. Later
on, the Hubble system has been refined and completed, introducing
spiral types later than Sc \citep[Sd and Sm;][]{deva59}, a new type of
amorphous galaxies \citep[Irr-II;][]{holm50} and ring-based
\citep{deva63} or arm-based \citep{elme87} distinctions among disk
galaxies.

Radically different classification schemes were proposed by
\citet{morg58} and \citet{vand59,vand60a,vand60b}. The first one links
morphology with central concentration of light and stellar
populations, also introducing the new cD type. The second one links
morphology with total luminosity (luminosity classes) and extends the
basic Hubble's scheme (E/Sp/Irr) to the lowest luminosity galaxies
(dwarfs).

Today the most frequently used classification scheme for statistical
studies is the numerical code associated to the so called {\it Revised
Hubble Type} ($T_{RH}$ hereafter), first introduced by \citet{deva74},
subsequently improved in the Third Reference Catalog of Bright
Galaxies \citep[][RC3 hereafter]{deva91} and schematically reminded
in Columns 1 and 2 of Table~\ref{RHT}.

For reasons which will become clear in the next Section, the MORPHOT
tool uses a slightly modified version of the $T_{RH}$ code (reported in
Column 3 of Table~\ref{RHT} as $T_M$: MORPHOT Type), in which the code -6
is associated to the cD galaxies (rather than to compact Es) and the
transition class E/S0 is introduced and coded as -4 (the code assigned
to cDs in the canonical $T_{RH}$ system).

\begin{table}
\begin{minipage}{85mm}
\caption{Revised Hubble Type ($T_{RH}$) and MORPHOT Type ($T_M$) codes}  
\label{RHT}      
\centering                          
\begin{tabular}{r l c}  
\hline\hline                 
Code & $T_{RH}$ & $T_M$ \\    
\hline                        
 -6 & cE$^*$ & cD \\
 -5 &  E & E \\
 -4 & cD & E/S0 \\
 -3 & S0$^-$ & S0$^-$ \\
 -2 & S0 & S0 \\
 -1 & S0$^+$ & S0$^+$ \\
  0 & S0/a & S0/a \\
  1 & Sa & Sa \\
  2 & Sab & Sab \\
  3 & Sb & Sb \\
  4 & Sbc & Sbc \\
  5 & Sc & Sc \\
  6 & Scd & Scd \\
  7 & Sd & Sd \\
  8 & Sdm & Sdm \\
  9 & Sm & Sm \\
 10 & Im & Im \\
 11 & cI$^*$ & cI$^*$ \\
\hline                                   
\end{tabular}
\end{minipage}
\\
\\
$^*$ cE and cI are the compact elliptical and irregular galaxies, respectively.
\end{table}

Before illustrating in some detail the MORPHOT tool, it is important
to explain what would be the ideal performance that we have tried to
achieve. This limit is obviously represented by the intrinsic
uncertainty of the morphological classifications provided by
experienced human classifiers.

In order to quantify this ideal target, we have first collected SDSS
g-band images of 163 galaxies in the redshift range 0.005-0.015, with
FWHM$\leq$2~arcsec and $T_{RH}$ classification code given in the RC3. In
this preliminar sample the fraction of early-type galaxies turned out
to be small compared with that in clusters. Therefore, we decided to
include in this sample 70 more SDSS images of early-type galaxies
obeying the same previous criteria about FWHM limit and morphological
availability on the RC3, but in the redshift range 0.015-0.03. It is
worth noting that, in spite of the worse image quality (average seeing
and photometric depth) of SDSS with respect to WINGS, due to the lower
range of redshift (z$\leq$0.03), most SDSS images of galaxies in our
sample have visual classification accuracy (spatial resolution in
kpc) at least comparable to that of the WINGS survey
(0.03$<$z$<$0.07; see Figure~8 in Paper~I).

   \begin{figure*}
   \centering
   \includegraphics[width=12.5truecm,angle=-90]{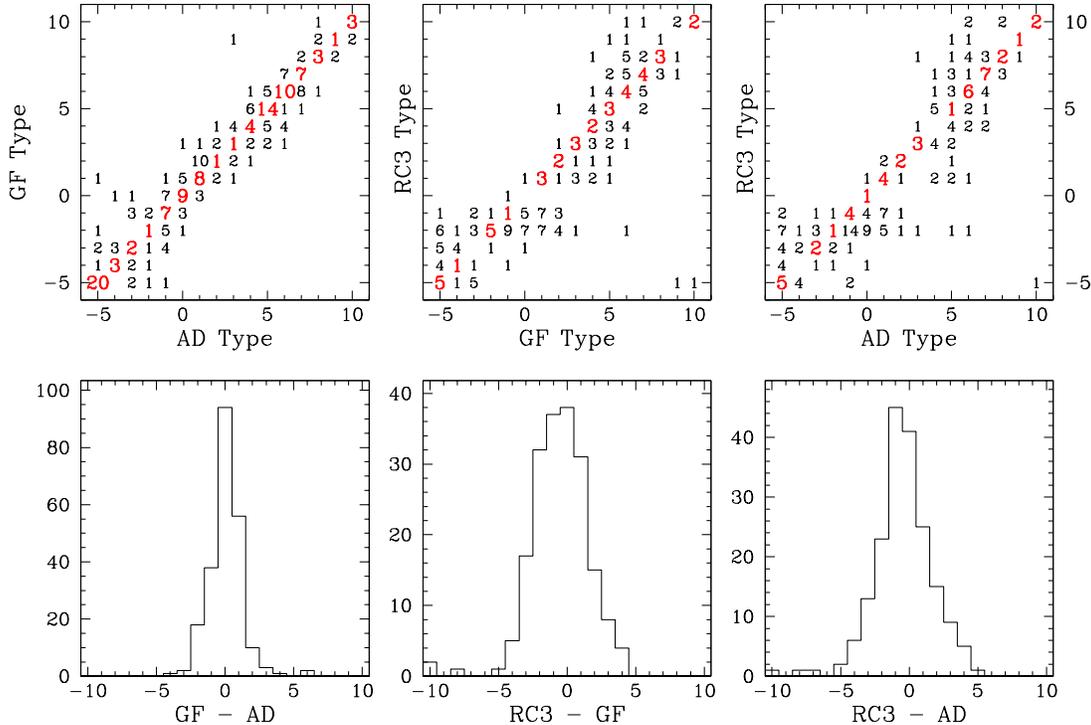}
      \caption{The top panels show the paired comparisons of the 
classifications from AD, GF and RC3, with the number of galaxies 
reported in each bin. The bottom panels report the histograms of 
the differences for each pair of classifiers}
         \label{VIScomp}
   \end{figure*}

The 233 galaxies in the final sample have been independently classified by
two of us (AD and GF) using the $T_{RH}$ code adopted in RC3. The top
panels of Figure~\ref{VIScomp} show the paired comparisons of the
classifications from AD, GF and RC3, with the number of galaxies
reported in each bin. Note that, since in the RC3 and AD databases
very uncertain classification has been assigned to 42 and 8 galaxies,
respectively (with 5 galaxies having uncertain morphology from both
RC3 and AD), the AD-GF, GF-RC3 and AD-RC3 comparisons just rely on
225, 191 and 188 common galaxies, respectively. Note also that compact
galaxies, both ellipticals ($T_{RH}$=-6) and irregulars ($T_{RH}$=11), are not
present in the selected sample of RC3 galaxies. The histograms of the
differences for each pair of classifiers are reported in the bottom
panels of Figure~\ref{VIScomp}, while the main statistical quantities
of these differences are reported in the first three rows of
Table~\ref{compvis}. 
{\bf The worse performances of the RC3 classifications with respect to those
given by AD end GF can be explained because of the very nature of the
RC3 data, which mainly result from compilation and statistical homogenization
of different (mostly inhomogeneous) data sources.}
The fourth and fifth rows of Table~\ref{compvis} report the same
quantities relative to the comparisons of the morphological type
estimates given by two of us (GF and WJC) for the clusters Abell~1643 and
Abell~1878 (z$\sim$0.20 and z$\sim$0.25, respectively; ground based,
very good seing imaging taken at the NOT) and for the clusters
CL~0024+16 and CL~0939+47 \citep[z$\sim$0.39 and z$\sim$0.41, respectively;
WFPC2 imaging from the MORPHS collaboration;][]{smai97}. These comparisons are
illustrated in \citet[][their Figure~2]{fasa00}. {\bf Comments about the last row
in the Table are given at the beginning of Section~\ref{secperf}.}
From Table~\ref{compvis} the visual morphological classifications turn
out not to be biased among each other, the largest average
displacement in the table being less than $\Delta T\sim$0.5. Instead, both the
$rms$ and the fractions of absolute differences less than 1, 2 and 3 times
$T_{RH}$ codes turn out to share relatively wide ranges ($\sigma_{\Delta T}$ from
1.2 to 2.4; $\vert\Delta$T$\vert\leq$1 from $\sim$0.53 to $\sim$0.84;
$\vert\Delta$T$\vert\leq$2 from $\sim$0.79 to $\sim$0.96; $\vert\Delta$T$\vert\leq$3
from $\sim$0.91 to $\sim$0.99). It is interesting to note that similar
uncertainties on the visual classifications and similar wide ranges in
the statistical quantities of the differences were found by the MORPHS
collaboration in their morphological catalog of 1857 cluster galaxies
at z$\sim$0.5, observed with WFPC2@HST and classified by 4 different
human classifiers \citep[][their Figure~1]{smai97}.

\begin{table*}
\caption{Comparisons among visual morphological classifications}             
\label{compvis}      
\centering                          
\begin{tabular}{c c c c c c c c c}  
\hline\hline                 
Comparison & $N_{gal}$ & z & Telescope & $<\Delta T>$ & $\sigma_{\Delta T}$ & $|\Delta T|\leq$1 & $|\Delta T|\leq$2 & $|\Delta T|\leq$3 \\    
\hline                        
 AD-GF  & 225 & $\leq$0.03 & SDSS &  0.076 & 1.257 & 0.836 & 0.960 & 0.982 \\
 GF-RC3 & 191 & $\leq$0.03 & SDSS & -0.554 & 2.374 & 0.529 & 0.796 & 0.932 \\
 AD-RC3 & 188 & $\leq$0.03 & SDSS & -0.425 & 2.272 & 0.569 & 0.787 & 0.910 \\
 GF-WJC &  67 & $\sim$0.2  & NOT  & -0.242 & 1.348 & 0.727 & 0.909 & 0.985 \\
 GF-WJC & 207 & $\sim$0.5  & HST  &  0.043 & 1.479 & 0.773 & 0.928 & 0.976 \\
 GF-GF & 136 & 0.04$\div$0.07  & INT+MPG  &  -0.072 & 1.158 & 0.940 & 0.976 & 0.994 \\
\hline                                   
\end{tabular}
\end{table*}

\section{The MORPHOT tool}\label{sectool}

\begin{figure*}
   \centering
   \includegraphics[width=15truecm]{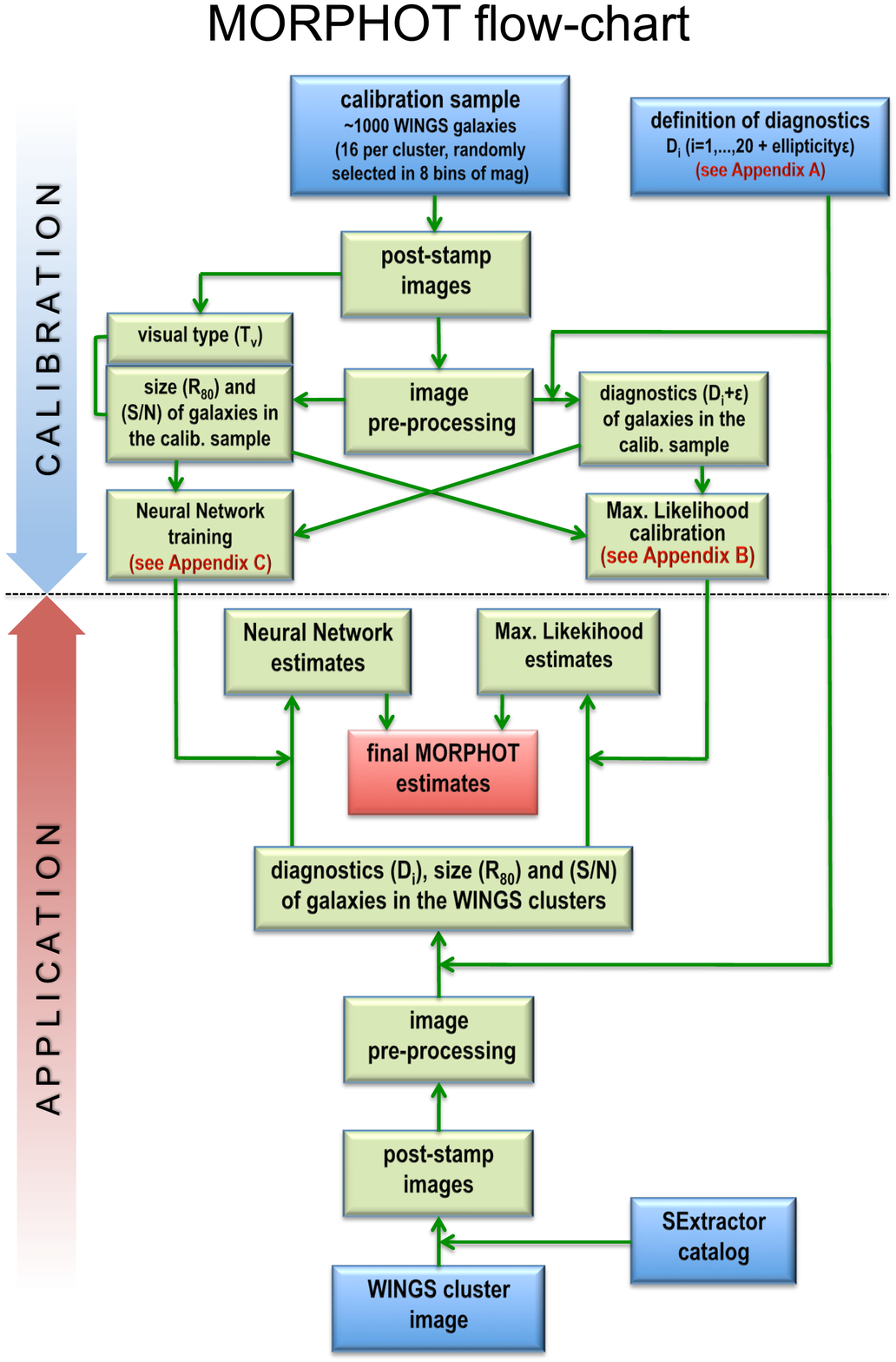}
      \caption{Flow-chart of MORPHOT}
         \label{FC1}
   \end{figure*}

Figure~\ref{FC1} shows the flow-chart of MORPHOT. The top and bottom
parts of the figure illustrate the calibration and application stages,
respectively. Each stage must be read following the direction of the big 
arrow on the left side. In particular, in the calibration stage, the visual
estimates $T_V$ (in the MORPHOT system $T_M$) are obtained for two
samples, each one including $\sim$1000 galaxies, extracted with the
same random ctiteria from the WINGS imaging. The first one will be
used as a calibration sample for the tool, while the second one will
be employed in Section~\ref{secperf} as a test sample in order to
assess the performances of MORPHOT. For each galaxy in the calibration
sample, (i) the global quantities: size (R), signal-to-noise ratio
($S/N$) and ellipticity ($\varepsilon$) are recorded; (ii) 20
image-based, numerical diagnostics of morphology ($D_i$, i=1,...,20)
are defined and their values are evaluated. The calibration sample is
used to gauge how the diagnostics $D_i$ depend on $T_V$ and on the
global quantities.  This allows us to produce a semi-analytical
estimator which combines the most effective diagnostics through a
Maximum Likelihood technique (ML; see Section~\ref{mltech} and
Appendix~\ref{appml}). The same sample is also used as a training set
for a Neural Network machine (NN; see Section~\ref{nntech} and Appendix~\ref{appnn}), in which
the global quantities (R, $S/N$ and $\varepsilon$) and the diagnostics
$D_i$ are the input quantities and the visual codes $T_V$ (in the
$T_M$ system of Table~\ref{RHT}) are the targets. Finally, the NN and
the ML estimators are combined to produce the final MORPHOT estimator
$T_M$.

In the following sub-sections of the present Section the various steps
of the MORPHOT tool are described in detail.

\subsection{The calibration sample of WINGS galaxies}\label{calsamp}

In the framework of the WINGS project, we have devised the
multi-object, automatic surface photometry tool GASPHOT
\citep{pign06,dono11}. This tool has been used to perform the detailed
surface photometry of {\bf 42297} galaxies in the WINGS clusters for which
SExtractor \citep{bear96} found more than {\bf 200~(300) contiguous pixels
(threshold area: A$_{thr}$) brighter than 1.25~(1.07) times the $rms$
per pixel of the background ($\sigma_{bkg}$) for those images obtained
with the WFC@INT~(WFI@ESO) \footnote{The different thresholds
  $\sigma_{bkg}$ and number of contiguous pixels take into account
  that we are using cameras with different pixelsize. See
  \citet{vare09} for details.}  ($\sim\mu_V$=25.7 for the
WINGS survey).}

With the aim of providing a sample of galaxies suitable
to calibrate MORPHOT, optimizing its performances for WINGS, we
decided to randomly extract 16 galaxies per cluster from the
WINGS--GASPHOT catalogs, taking care of putting two galaxies
in each one of the 8 bins of apparent V magnitude defined as follows:
V$\leq$15, 15$<$V$\leq$16, ... , 20$<$V$\leq$21, V$>$21. In this way
we gathered 1216 WINGS galaxies sampling uniformly the whole range of
magnitudes of the GASPHOT--WINGS galaxy sample (see
Section~\ref{secperf}) and spanning the whole range of observing conditions
(background noise and FWHM) of the WINGS optical imaging
\citep[see][]{vare09}. We decided to remove from this sample those
galaxies too close to the edges of the CCDs and/or the very peculiar
objects (on-going mergers or quite ill-shapen galaxies). After that,
we are left with a final calibration sample of 926 galaxies. All these
galaxies have been visually classified by GF according to the $T_M$ code
shown in Table~\ref{RHT}.

It is worth recalling here that the photometry of the WINGS optical
survey has been performed on images in which large galaxies and halos
of bright stars have been removed after modeling them with elliptical
isophotes \citep[see][]{vare09}. This allowed us to perform a careful
subtraction of the background and estimation of its $rms$
($\sigma_{bkg}$) even for small galaxies embedded in the halo of the
brightest cluster galaxies. Therefore, the WINGS optical catalogs
provide, for all galaxies, robust SExtractor determination of the
ellipticity ($\varepsilon_{SEx}$) and of the above mentioned threshold
area A$_{thr}$.

It is also worth mentioning that each individual galaxy is recorded
within a square, odd-sized frame of side 3$\times$a$_{maj}$, where
a$_{maj}$ corresponds to the semi-major axis (in pixels) of the
ellipse with area A$_{thr}$ and ellipticity $\varepsilon_{SEx}$.

\subsubsection{Preliminar image processing}\label{preproc}

Before running the core-tool of morphological type estimation, for
each post-stamp galaxy image of a given sample, MORPHOT automatically
performs the refinement of the local background subtraction and the
galaxy re-centering. In particular, the central pixel of each galaxy
image is made to be coincident with the intensity peak or with the
distance-averaged intensity (bari)center, depending on whether the
galaxy shows a well defined, dominant light peak (regular shape) or an
irregular structure with several local peaks. Moreover, a preliminar
processing of the post-stamp galaxy image is performed, which produces
two ancillary (and temporary) frames $F_C$ and $F_S$. 

In the frame $F_C$ (C:~Clean), the possible spurious features (ghosts)
and/or those objects (both stars and galaxies) different from the
galaxy under analysis are removed by comparing each other the original
($F_0$)and the 180$^\circ$ rotated ($F_{180}$) images. In particular,
if at a given pixel position the difference $F_0$-$F_{180}$ is greater
than $n$ times the $rms$ of the pixel values of $F_{180}$ over a box
of side = 2$\times$FWHM around the same position, the pixel value in
$F_0$ is replaced by the corresponding value in $F_{180}$. We have
empirically verified that, in our redshift range and with our
instrumental set, using $n$=3 allows us to remove satisfactorily most of
the unwanted objects without either changing the statistical
properties of $F_0$ (image texture), or fading the interesting galaxy
features, like spiral arms, bars, rings and HII regions.

The frame $F_S$ (S:~Smooth/Symmetric) is obtained in two steps: first
the symmetrization is achieved by averaging (pixel by pixel) $F_C$
with its 180$^\circ$ rotated version; then, the median (3$\times$3)
and adaptive \citep[][max. block size=11]{rich91} filters are applied
to the symmetrized frame.

It is worth stressing that the evaluation of the morphological
diagnostics (see following Sections) is performed on either $F_C$ or
$F_S$, depending on the particular diagnostic. In general, those more
specifically linked to the global properties of galaxies (i.e.:
different kinds of concentration, etc..)  are evaluated on $F_S$,
while those dealing with pixel-scale structures, local features
(clumpiness, diskyness, etc..) and simmetry are evaluated on $F_C$.
The cleaned image $F_C$ is also used to determine the total intensity
(I$_T$, in ADUs) and the final, global geometrical parameters
(ellipticity $\varepsilon$ and position angle $\theta$) of the
galaxy. These are used in turn to produce a model image $F_M$ from the
elliptical apertures intensity profile of $F_C$, to determine the
equivalent radii (in pixels) enclosing 80\% of the total galaxy light
(R$_{80}$) and to compute the average signal-to-noise ratio of the
galaxy: $S/N$=I$_T$/($\sigma_{bkg}$A$_{thr}$).  The global quantities
$\varepsilon$, R$_{80}$ and $S/N$ are used in the calibration
procedure of diagnostics (see Section~\ref{secdef}).

\subsection{Morphological diagnostics}\label{secdef}

Our approach to the automation of morphological classification is a
fully empirical one. We do not try to identify an orthogonal set of
few, independent and physically pregnant morphological indicators
(diagnostics, hereafter), as in the case of the CAS parameter set
\citep[Concentration, Asymmetry and Clumpiness;][]{cons03} or in the
papers by \citet{vand08} and \citet{scar07}. Rather, we prefer to bet
on a large number of diagnostics, no matter if in some cases they are
similar among each other, since we postulate that each one of them
could potentially be sensitive to some particular morphological
characteristic and/or feature of the galaxies. In other words, we
decided not to throw out anything a priori and to defer to a later
stage the possiblity of giving up some of the diagnostics. As well, we
do not try to select the most significant diagnostics by means of
statistical techniques, like for instance the Principal Component Analysis (PCA, see
Section~\ref{seccomb}). We just test each diagnostic on the field (the
test sample, see Section~\ref{secperf}), checking whether its addition
to the previous (smaller) set of diagnostics improves the tool's
performances (see Section~\ref{mltech} and Figure~\ref{dialoop}). 
Lastly, our diagnostics are not necessarily defined (and
conceived) to be independent of the image parameters (photometric depth,
noise, pixel size, seeing, etc..). In fact, at least for the time
being, we aim to apply MORPHOT just to the WINGS imaging, deferring to
a later time the release of a more generally usable version of the
tool, where the definition of the diagnostics will be as most as
possible independent on the observing material.  More explicitly, the
calibration of the diagnostics we describe in the next Sections is
performed on the WINGS calibration sample defined in
Section~\ref{calsamp} and holds good just for WINGS-like data.  For
now, applying MORPHOT to imaging data different from WINGS would
actually imply a re-calibration of the diagnostics on the new data
set.

\begin{figure*}
  \vspace{-2.0truecm}
  \includegraphics[width=19truecm]{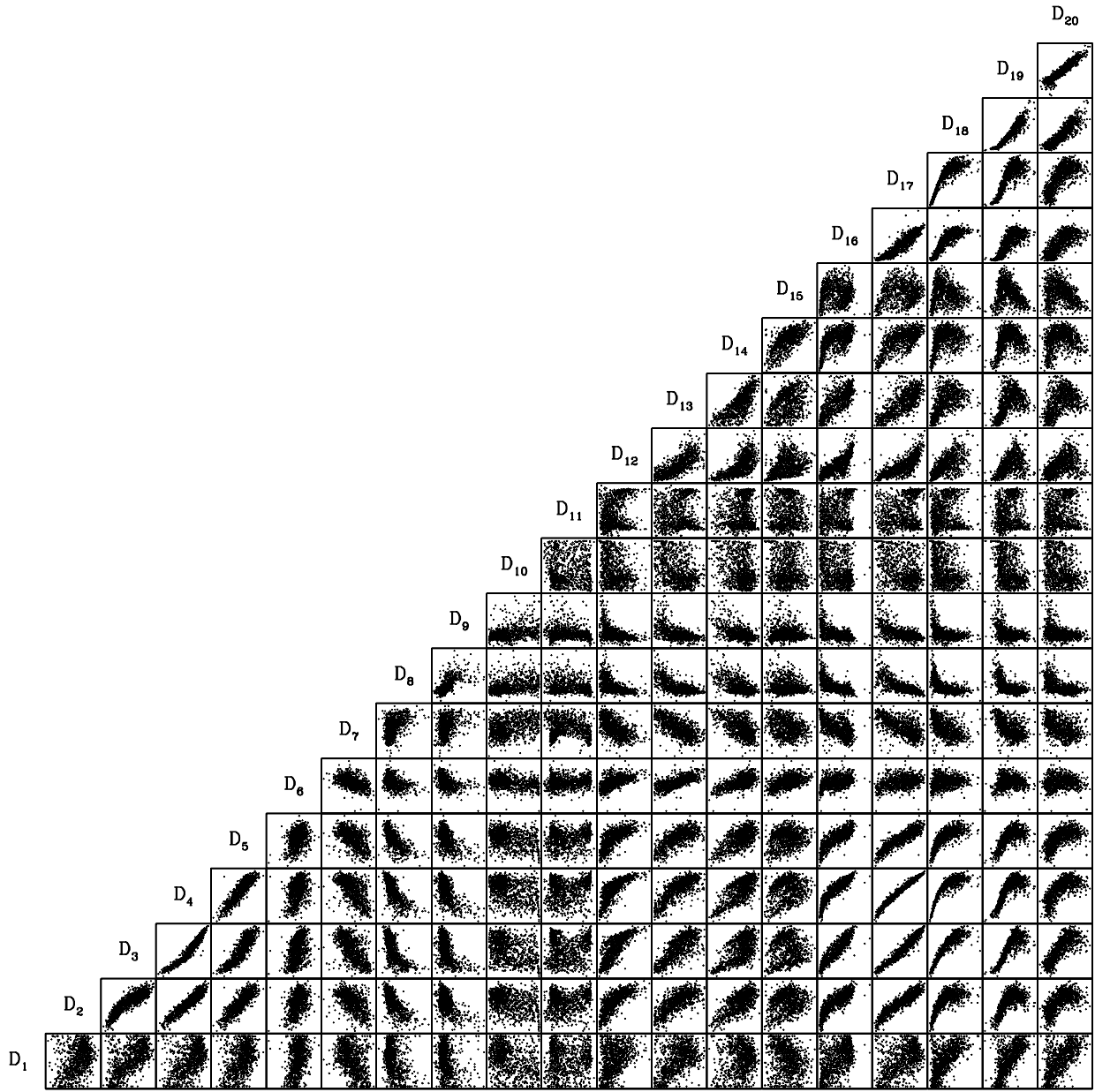}
   \vspace{-5.0truecm}
   \caption{Mutual correlations among the MORPHOT diagnostics {\bf defined in Appendix~\ref{appdia}}.}
   \label{parcorr}
  \end{figure*}

Up to now we have empirically introduced (and tested) 20 diagnostics.
Some of them are not conceptually original, but are usually more
simply defined with respect to the similar indicators already present
in the literature. Again, we prefer to test a large number of rough
diagnostics that are rapidly evaluated, rather than a small set of carefully
calibrated (but sometimes hard to compute and not necessarily more
efficient) indicators. It is also worth noting that our set of
diagnostics is actually open, meaning that additionally devised
diagnostics \citep[like, for instance, the spirality analyzers from][]{naim95,sham11} 
can be introduced without changing the structure of the
tool. The only limitation we pose for the new diagnostics is that they
have to be image-based, thus excluding color- and spectroscopy-based
quantities. This is because we think that, in order to have an
unbiased picture of the evolution of galaxies in clusters, the
information on morphology and stellar population should be kept
separate. In Appendix~\ref{appdia} we present in some
detail the definition and the meaning of the 20 diagnostics $D_i$
(i=1,...,20). Here we just mention that many of them turn out to be
correlated, sometimes strongly so. This is shown in
Figure~\ref{parcorr} for the calibration sample 
and it is somehow expected due to our empirical approach.

\subsubsection{Diagnostic dependences}\label{secdep}

Figure~\ref{parT} illustrates how the 20 morphological diagnostics
{\bf defined in Appendix~\ref{appdia}} correlate with the visual
morphological type for the 926 galaxies of the calibration sample. For
the sake of clarity the Y-labels are omitted in the figure and the
visual morphological types (integer values) of the calibration
galaxies in the southern clusters (304 objects) are shifted by 0.5
upward.

  \begin{figure}
  \vspace{-2.0truecm}
  \hspace{-1truecm}
  \includegraphics[width=10.5truecm]{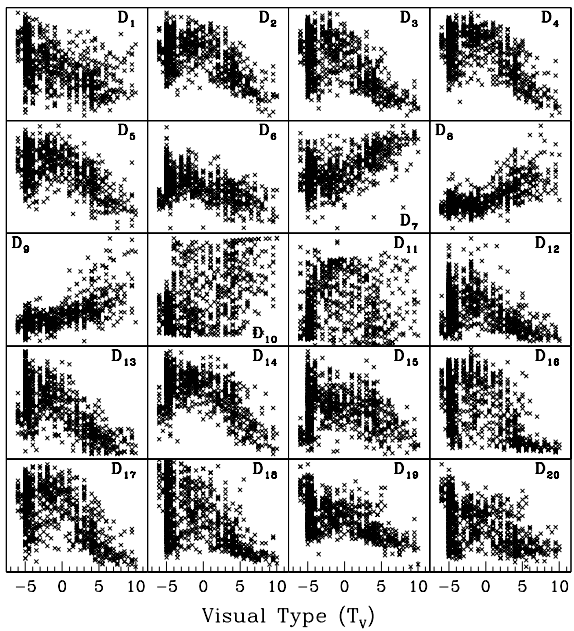}
   \vspace{-3.0truecm}
   \caption{The MORPHOT diagnostics {\bf defined in Appendix~\ref{appdia}} vs. visual morphological type {\bf for the calibration sample}.}
   \label{parT}
  \end{figure}

\begin{figure}
   \vspace*{-1.5truecm}
   \centering
   \hspace*{-1truecm}
   \includegraphics[width=10truecm]{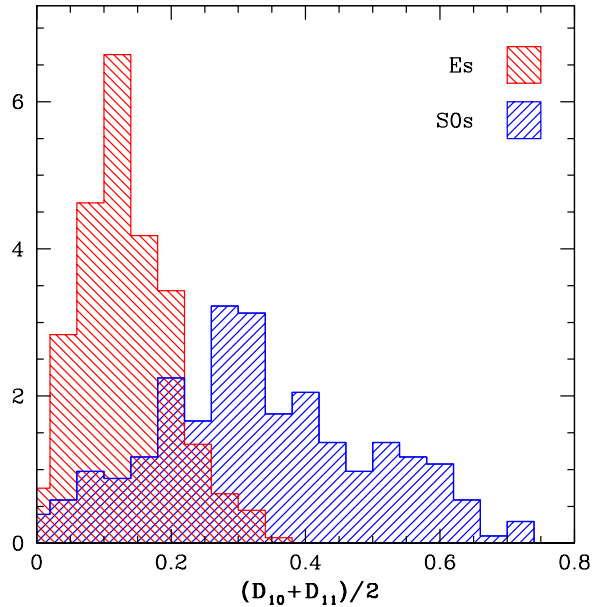}
   \vspace*{-3truecm}
      \caption{Distribution of the average value of the diskiness diagnostics
$D_{10}$ and $D_{11}$ {\bf (see Appendix~\ref{appdia})} for the (visually classified) elliptical and 
S0 galaxies of the calibration sample}
         \label{diskES0}
   \end{figure}

 Moreover, in Figure~\ref{parT} the diagnostics $D_1$, $D_7$, $D_8$,
$D_9$, $D_{10}$, $D_{12}$, $D_{15}$, $D_{19}$ and $D_{20}$ are plotted
in logarithmic scale. It is evident from the figure that many
diagnostics have quite similar (average) behaviour as a function of
the visual morphological type. Still, as explained before, we
postulate that even slight differences among diagnostics could in
principle help to disentangle different morphological features and we
defer the final decision about the diagnostics to be retained to the
comparison of the results with the visual classifications. For the
moment, it is worth emphasizing the importance of the diagnostics
$D_{10}$ and $D_{11}$ in disentangling ellipticals from S0
galaxies. Figure~\ref{diskES0} shows that the distributions of these
diagnostics for the two morphological types are quite apart. The
relative scarcity of non--disky (face-on) S0s in the figure is likely
attributable to some cases of face-on S0s which have been visually
misclassified as ellipticals. To this concern we note that, with the
spatial resolution typical of the WINGS survey, to entirely remove
this kind of miss-classification turns out to be almost impossible,
even for visual classifications. Actually, \citet{capa91} have shown
that to distinguish face-on S0s from ellipticals could be a difficult
task even for very well resolved galaxies (see \citealp{capp11} for a
more radical point of view).

\begin{figure*}
 \vspace{-1.5truecm}
 \centering
 \hspace{-1truecm}
 \includegraphics[width=10.5truecm]{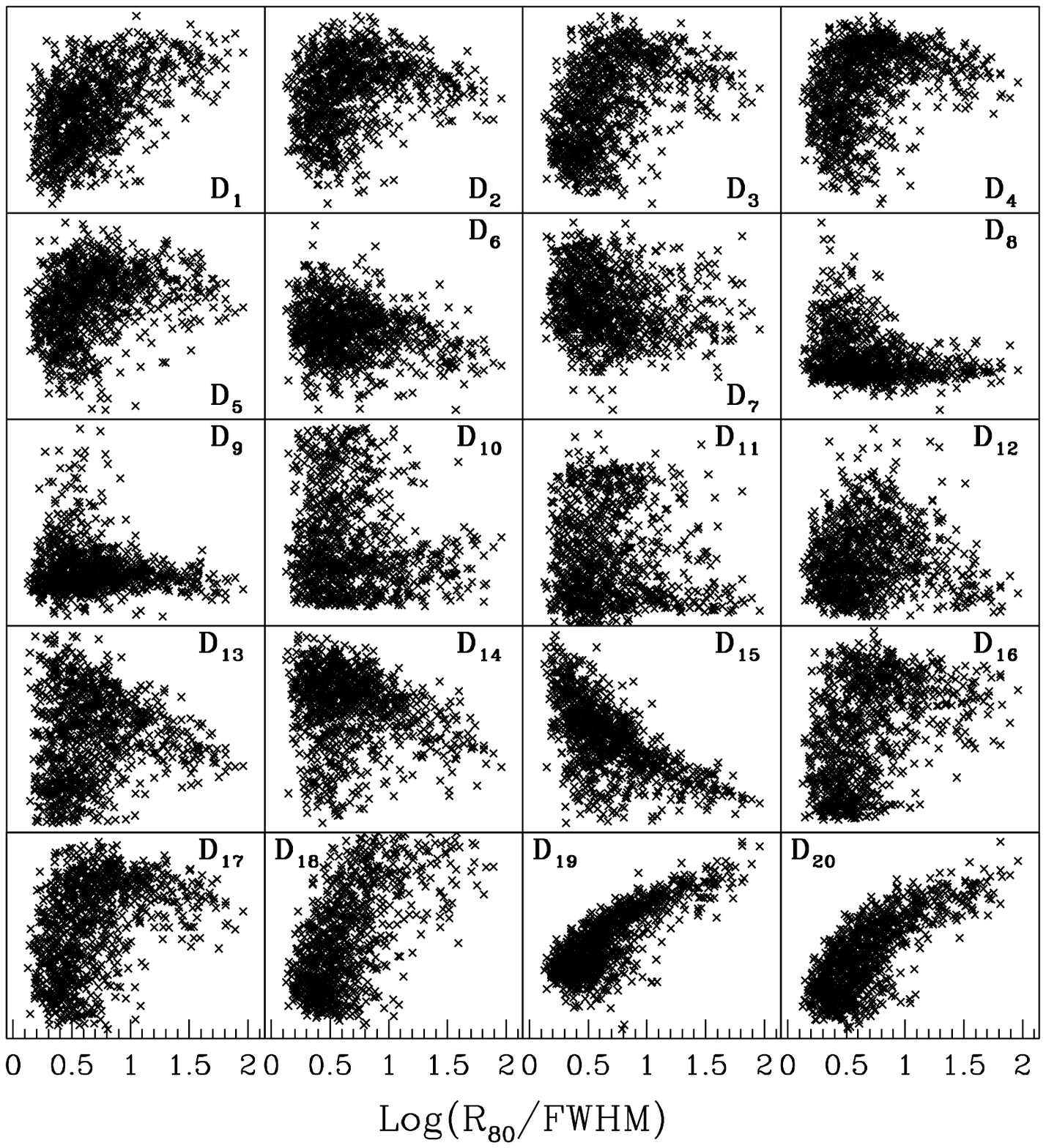}
 \hspace{-3truecm}
 \includegraphics[width=10.5truecm,origin=c]{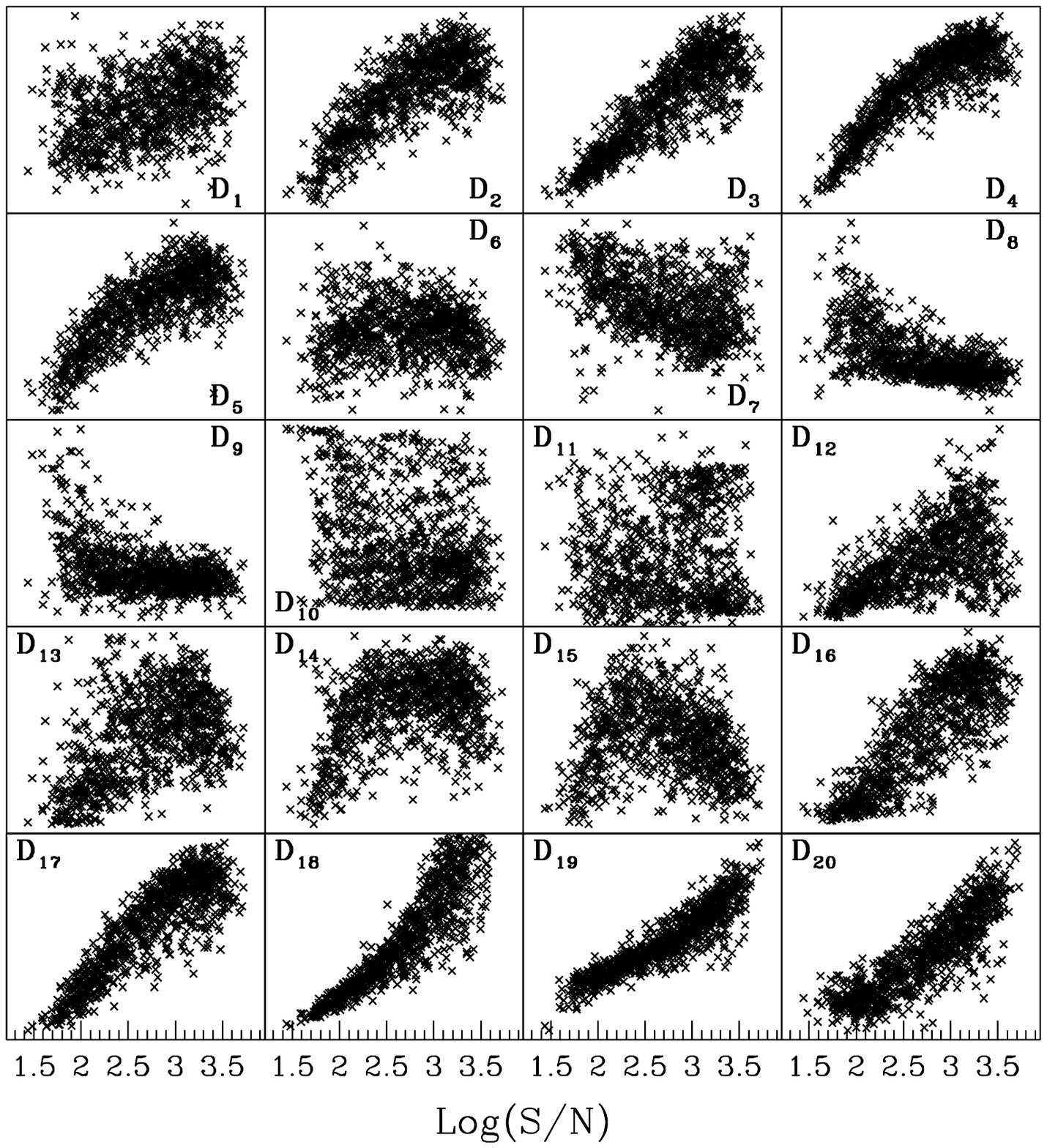}
  \vspace{-3.0truecm}
  \caption{The MORPHOT diagnostics {\bf defined in Appendix~\ref{appdia}} vs. $\log(R_{80}/FWHM)$ and $\log(S/N)$.}
   \label{parRSN}
  \end{figure*}

As already stated before, since our aim is to provide morphological
classifications of WINGS galaxies, our diagnostics $D_i$ are not
conceived to be independent on instrumental and observing parameters
(pixelsize, seeing, $S/N$, etc..). Figure~\ref{parRSN}
illustrates how the $D_i$ depend on $\log(R_{80}/FWHM)$ and on
$\log(S/N)$.  
 
Instead, no significant dependence of the diagnostics on
the apparent ellipticity $\varepsilon$ has been found.  To this
concern, before describing the techniques we used to extract from our
diagnostics univocal estimates of the morphological type, it is
worth mentioning that, from now on, we formally include $\varepsilon$
into the set of diagnostics.  Therefore, their total number is
hereafter assumed to be of 21 ($D_i$,i=1,..,20+$\varepsilon$).

\subsection{Combining the diagnostics}\label{seccomb}

Having defined the diagnostics $D_i$ and tested their dependence
on both the visual morphological type ($T_V$) and the global
quantities R$_{80}$ and $S/N$, we are left with the difficult task of
simultaneously exploiting their capabilities, in order to improve as
much as possible the final effectiveness of the tool in recognizing
the morphology of galaxies. In other words, we must combine in some
(smart) way the 21 diagnostics to obtain a single, final morphological
estimator.

Although our empirical approach would drive us to use all the
diagnostics (see Section~\ref{secdef}), we first tried to identify,
through the canonical Principal Component Analysis (PCA), an
orthogonal transformation converting our diagnostics in a set of
uncorrelated variables, smaller than the original set, but still
preserving the wealth of morphological information contained
therein. However, likely because our diagnostics are not normally
distributed, this attempt turned out to be unsuccessful. In fact,
running the PCA on our galaxy calibration sample, we just obtained two
significant eigenvectors, whose linear combination resulted in an
extremely large scatter of the PCA morphological types with respect to
the visual estimates. Actually, in Section~\ref{mltech} (see
Figure~\ref{dialoop}) the number of significant diagnostics is shown
to be much larger than two.

Returning to our empirical approach, we used two different techniques,
totally independent on each other, in order to obtain the above
mentioned combination of the diagnostics and the final, global
morphological estimator. It is worth mentioning that both techniques
produce morphological type estimates in one digit decimal numbers.

\subsubsection{Maximum Likelihood estimator}\label{mltech}

As outlined at the beginning of Section~\ref{sectool} (see also the
flow-chart in Figure~\ref{FC2}), the first technique exploits the
Maximum Likelihood (ML) statistics to combine the diagnostics.
Concisely, after having removed their obvious dependences on the galaxy
size (relative to the FWHM) and signal to noise ratio (see
Section~\ref{secdep} and Figure~\ref{parRSN}), we use the dependence of diagnostics on the
visual morphological type in the calibration sample (i.e. the 2D
distributions in Figure~\ref{parT} to
estimate the probability that a given value of each diagnostic could
come from (be measured for) galaxies of all possible morphological
types. Then, for a galaxy with unknown morphology and known (measured)
diagnostics, we compute the ML probability (product of the
probabilities associated to the diagnostics) as a function of the
morphological type $T$ and we assume that the '{\it true}'
morphological type of the galaxy is that providing the largest value
of ML.  From the function ML($T$), we can also derive the confidence
interval of each ML estimate of the morphological type. Details about
the MORPHOT-ML technique can be found in Appendix~\ref{appml}.

\begin{figure}
   \vspace*{-1.5truecm}
   \centering
   \hspace*{-1truecm}
   \includegraphics[width=10truecm]{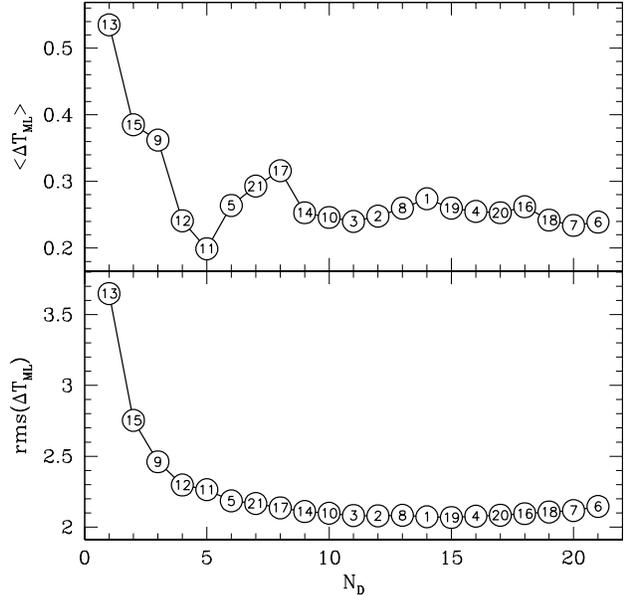}
   \vspace*{-3truecm}
      \caption{Average value (upper panel) and $rms$ (lower panel) of the differences
$\Delta T_{ML}$=($T_{ML}-T_V$) as a function of the number of diagnostics
($N_D$) used for the ML estimator. The numbers in the open circles identify
the diagnostics (see Appendix~\ref{appdia}) recursively added to the previus 
ones. The average value and the $rms$ become nearly stable after $N_D$=9
and $N_D$=11, respectively}
         \label{dialoop}
   \end{figure}

In order to determine how many diagnostics (and which ones) are
necessary (and sufficient) to optimize the ML techique, we have first
applied the above procedure to all galaxies in the calibration sample
using the 21 $D_i$ one by one and recording the diagnostic which
provides the lowest scatter ($rms$) of the differences between the ML
and visual morphological types ($T_{ML}-T_V$; hereafter $\Delta
T_{ML}$). Then, we have repeated the procedure by adding the remaining 20
diagnostics one by one to the firstly selected diagnostic, and we have
again recorded the one which minimizes the above mentioned $rms$ among
the 20 couples of diagnostics. We iterated this loop, each time adding
one by one the remaining $n$ diagnostics to the (21-$n$) already
recorded, while the $D_i$ last. Figure~\ref{dialoop} illustrates
the result of this iteration, showing how the average value
and the $rms$ of the $\Delta T_{ML}$ distribution vary as a function of the
number $N_D$ of diagnostics used to provide the $T_{ML}$ of galaxies in the
calibration sample.

The circled numbers in the figure refer to the corresponding
diagnostic's numbers in Appendix~\ref{appdia}. Figure~\ref{dialoop}
shows that the average $\Delta T_{ML}$ becomes nearly stable (at
$\sim$0.24) after $N_D\sim$9, while the $rms$ of the $\Delta T_{ML}$
distribution decreases till $N_D\sim$11. It is worth mentioning that,
according to the $F$-test for significantly different variances, up to
this value of $N_D$, the addition of new diagnostics significantly
reduces the $rms$, at variance with the (above mentioned) formal
result of the PCA. From Figure~\ref{dialoop} the diagnostics which
turn out to be effective for the ML technique, sorted by decreasing
effectiveness, are: $D_{13}$, $D_{15}$, $D_9$, $D_{12}$, $D_{11}$, 
$D_5$, $\varepsilon$, $D_{17}$, $D_{14}$, $D_{10}$ and $D_3$.

The outlined iterative procedure, aimed at identifying the most
effective diagnostics of the ML technique, is clearly an empirical
one. {\bf For instance, we have chosen to stop the iterations when
minimizing the $rms$ ($N_D$=11), rather than the average value of
$\Delta T_{ML}$, since we consider the minimum $\Delta T_{ML}$ at
$N_D$=5 to be just a statistical fluctuation due to the finiteness of
the test sample}. Moreover, we cannot rule out (actually, we consider
very likely) the possibility that different combinations of $D_i$ coud
work better, giving lower values of $rms(\Delta T_{ML})$. Still,
testing all possible combinations of the diagnostics was intractable
and, we believe, unproductive. Thus, we assume that the morphological
types $T_{ML}$ we got using the first eleven effective diagnostics are
the best possible ML estimates for the calibration galaxies. {\bf Although
lacking in a rigorous explanation, we believe the slightly worse $rms$
performance of the tool for $N_D>$11 to be due (again) to the
finiteness of the test sample, which might induce in the empirical ML
procedure a sort of oversampling noise.}

The short-dashed lines in Figure~\ref{prest} connect the median values of $\Delta T_{ML}$ in
different bins of $T_V$ (upper panel) and the corresponding $rms$
values (lower panel), while the relevant global statistics of the 
$\Delta T_{ML}$ distribution for the calibration sample are reported
in the first row of Table~\ref{compvis1}.

\subsubsection{Neural Network estimator}\label{nntech}

The second technique we use to combine the morphological diagnostics
is based on the classical feed-forward multilayer perceptron Neural
Network (NN). Details about the MORPHOT-NN technique can be found in
Appendix~\ref{appnn}. Here we just mention that again the NN
morphological type estimates are supplied with confidence intervals,
while in this case (at variance with the ML technique) we use as input
quantities of the NN machine the whole set of diagnostics ($D_i$,
i=1,...,20) plus the global quantities $\varepsilon$,
$\log(R_{80}/FWHM)$ and $\log(S/N)$. The reason for this choice is
explained in Appendix~\ref{appnn}.

The dot-dashed lines in Figure~\ref{prest} (blue in the electronic
version) connect the median values of $\Delta T_{NN}$ in different
bins of $T_V$ (upper panel) and the corresponding $rms$ values (lower
panel), while the relevant global statistics of the $\Delta T_{NN}$
distribution for the calibration sample are reported in the second row
of Table~\ref{compvis1}.  Comparing these values with the
corresponding ones relative to the ML technique (first row of the same
table) and looking at Figure~\ref{prest}, we conclude that the
performances of the NN estimator are significantly better than those
of the ML estimator.

\subsubsection{The final MORPHOT estimator}\label{secfinest}

  \begin{figure}
   \centering
   \hspace*{-1truecm}
   \includegraphics[width=10.5truecm]{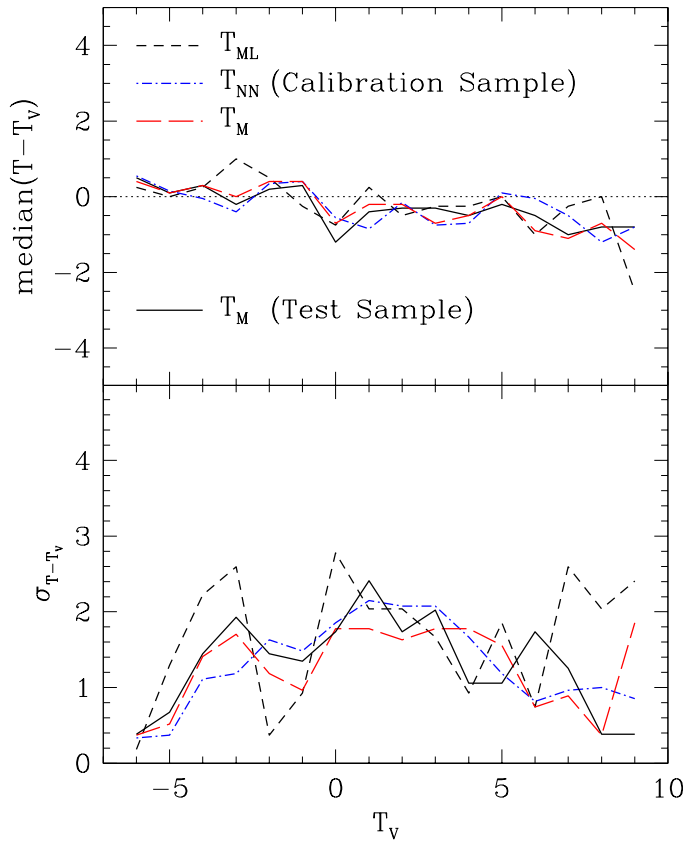}
     \caption{{\it Upper panel}: The median values of ($T-T_V$) in
different bins of $T_V$ are connected for the estimators $T_{ML}$
(short-dashed line), $T_{NN}$
(dot-dashed line, blue in the electronic version) and $T_M$
(long-dashed line, red in the electronic version) obtained running
MORPHOT on the calibration sample. The full (black) line 
illustrates the behaviour of the $T_M$ estimators obtained running
MORPHOT on the test sample (see Section~\ref{secperf}).
{\it Lower panel}: The $rms$ of ($T-T_V$) in different bins of $T_V$ 
for the estimators $T_{ML}$, $T_{NN}$ and $T_M$ in the case of the
calibration sample. The meaning of the different lines are as in the
upper panel. The full (black) line refers to the $T_M$ estimator in
the case of the test sample.}
         \label{prest}
   \end{figure}

As already pointed out at the beginning of Section~\ref{seccomb}, the
Maximum Likelihood and the Neural Network provide conceptually and
technically different approaches to the problem of combining the
morphological diagnostics. Therefore, the MORPHOT estimators produced
by the two techniques ($T_{ML}$ and $T_{NN}$) should be independent on
each other and the $rms$ of their difference should roughly be the
square root of the sum of their variances with respect to $T_V$
($rms[T_{ML}-T_{NN}]\sim$2.55). Actually, elementary numerical
simulations show that the particular density distribution of $T_V$
makes the real $rms$ lower than the above theoretical value,
in agreement with the value we found in the calibration sample 
($rms$=2.05, see last row in Table~\ref{compvis1}).

Once the mutual independence of the two estimators has been checked,
the last step of the MORPHOT flow-chart (see Figure~\ref{FC1}) is the
evaluation of the final morphological type estimator $T_M$, which is
simply defined as the average value of the two independent estimators:
$T_M$=($T_{ML}+T_{NN}$)/2. Similarly, the lower and upper limits of
the confidence interval of $T_M$ are obtained by averaging the lower
and upper confidence limits of $T_{ML}$ and $T_{NN}$, respectively.

The long-dashed lines in Figure~\ref{prest} (red in the electronic
version) connect the median values of $\Delta T_M$ in different bins
of $T_V$ (upper panel) and the corresponding $rms$ values (lower
panel), while the relevant global statistics of the $\Delta T_M$
distribution for the calibration sample are reported in the third row
of Table~\ref{compvis1}.

Comparing these values with the corresponding ones relative to the ML
and NN techniques (first two rows of the same table), we should
conclude that the global performances of the NN estimator are even
better than those of the final MORPHOT estimator, thus making
convenient to adopt $T_{NN}$ alone to optimize the
performances of MORPHOT. Still, just because of the above mentioned
mutual independence of the ML and NN estimators, we are inclined to
believe that their combination could in any case provide more stable
results, each technique possibly compensating the biases of the other
one. Actually, from the upper panel of Figure~\ref{prest}, the biases
of the $T_M$ and $T_{NN}$ estimators in different bins of $T_V$
have quite similar sizes, while the scatter of the $T_M$
estimator along the $T_V$ sequence (long-dashed, red line in the lower panel) 
turns out to be more stable than in the case of the $T_{NN}$ estimator
(dot-dashed, blue line in the same panel). Therefore,
we decided to adopt $T_M$ as final MORPHOT estimator.

It is worth noting that all the estimators tend to be biased towards later
and earlier morphological types for the early and late visual types, respectively.
However, this is expected (and in some sense obvious) because of the 
one-sided error distribution of galaxies close to the limits of the
available range of visual morphological types. 

\section{Testing the performances of MORPHOT}\label{secperf}

As mentioned in Section~\ref{sectool}, the test galaxy sample has been
extracted from the V-band WINGS imaging using the same (random)
criteria described in Section~\ref{calsamp} for the calibration
sample. Again, we removed from the initial sample of 1216 galaxies
those objects too close to the edges of the CCDs and/or the very
peculiar objects (on-going mergers or quite ill-shapen galaxies),
being left with a final sample of {\bf 979} objects, which have been
visually classified by GF according to the $T_M$ code exemplified in
Table~\ref{RHT}. It is worth noting that {\bf 136} (mostly bright) galaxies
turned out to be in common between the calibration and the test
samples. Since the classifications of the two samples (both from GF)
are independent from each other, these galaxies have been used to
estimate the internal consistency of the visual classifications from
GF. {\bf The main statistical indicators of this comparison are reported in
the last row of Table~\ref{compvis}. They show that significant
differences can be found even comparing among each other the 
morphological classifications provided by the same human classifier, 
for the same galaxy sample, but at different times.} 

For all the galaxies in the test sample, we have computed the global
quantities ($\varepsilon$, $R_{80}$ and $S/N$) and the MORPHOT
diagnostics $D_i$ and we have obtained the Maximum Likelihood
($T_{ML}$), the Neural Network ($T_{NN}$) and the final ($T_M$)
MORPHOT estimators of the morphological type.

\begin{table*}
\caption{Comparisons between visual morphology and MORPHOT Type 
for both the calibration and the test samples. The last row refers
to the mutual comparison between $T_{ML}$ and $T_{NN}$.}             
\label{compvis1}      
\centering                          
   \hspace*{-1truecm}
\begin{tabular}{c c r c c c c c}  
\hline\hline                 
$\Delta T$ & Sample & $<\Delta T>$ & $\sigma_{\Delta T}$ & median($\Delta T$) & $|\Delta T|\leq$1 & $|\Delta T|\leq$2 & $|\Delta T|\leq$3 \\    
\hline                        
 $T_{ML}-T_V$ & CALIB & 0.24& 2.08 & 0.00 & 0.616 & 0.778 & 0.867 \\
 $T_{NN}-T_V$ & CALIB & -0.05 & 1.47 & 0.10 & 0.656 & 0.875 & 0.951 \\
 $T_M-T_V$ & CALIB & 0.01 & 1.56 & 0.10  & 0.640 & 0.832 & 0.925 \\
 $T_M-T_V$ & TEST & -0.06 & 1.72 & 0.00 & 0.631 & 0.803 & 0.885 \\
\hline                                   
 $T_{ML}-T_{NN}$ & CALIB & 0.29& 2.05 & 0.10 & 0.522 & 0.754 & 0.865 \\
\hline                                   
\end{tabular}
\end{table*}

In {\bf the fourth row of} Table~\ref{compvis1} we report the relevant statistical quantities
of the comparisons between visual and MORPHOT classifications for the test sample.

From Tables~\ref{compvis} (rows 1, 4 and 5) and \ref{compvis1} (rows 3
and 4) and from Figure~\ref{prest} one can derive the
following remarkable conclusions: (i) for the calibration sample the
scatter of the $T_M$ estimator with respect to the visual
classifications turns out to be quite comparable to (sometimes better
than) the scatter reported in Table~\ref{compvis} among visual types
provided by different experienced human classifiers; (ii) for the test galaxy
sample, the above mentioned scatter just marginally increases with
respect to the previous case, still remaining quite competitive with
respect to the average scatter among visual classifications.

 \begin{figure}
   \centering
   \hspace*{-0.5truecm}
   \includegraphics[width=9.5truecm]{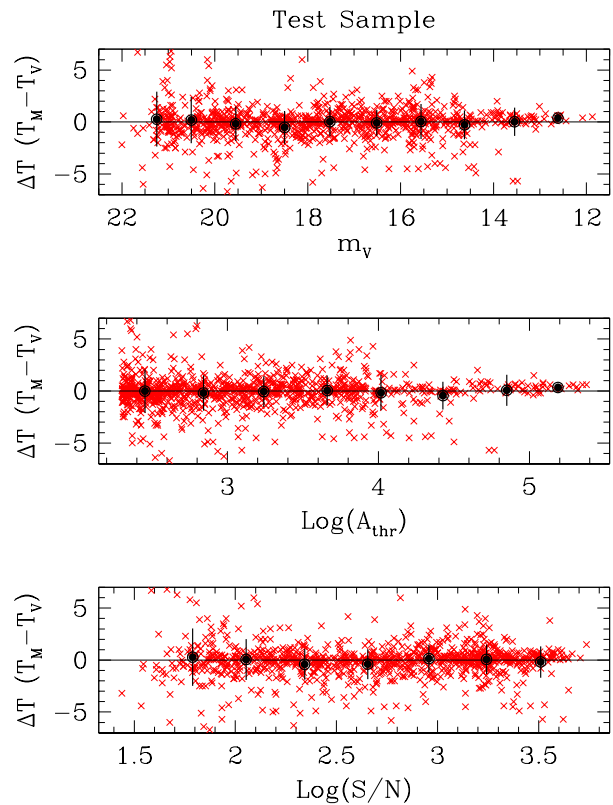}
     \caption{Dependence of the difference ($T_{M}-T_V$) 
on apparent magnitude (upper panel), threshold area (middle panel) and 
signal to noise ratio (bottom panel) for the test sample. {\bf The big, full dots
and the corresponding error bars illustrate the average values ond the $rms$
in each bin.}}
         \label{TFdep}
   \end{figure}

Figure~\ref{TFdep} illustrates how the differences $\Delta
T_M$=$T_M$-$T_V$ (average value and scatter) behave as a function of
apparent magnitudes, threshold areas and the S/N ratios of galaxies in
the test sample. As might be expected, the scatter of $\Delta T_M$ increases 
(from $\sim$0.7 to $\sim$2) at decreasing
the threshold area (in pixels) and the apparent luminosity of galaxies.
Instead, no dependence of the scatter on $S/N$ is found.

  \begin{figure}
   \vspace*{-0.5truecm}
   \centering
   \hspace*{-1truecm}
   \includegraphics[width=10.5truecm]{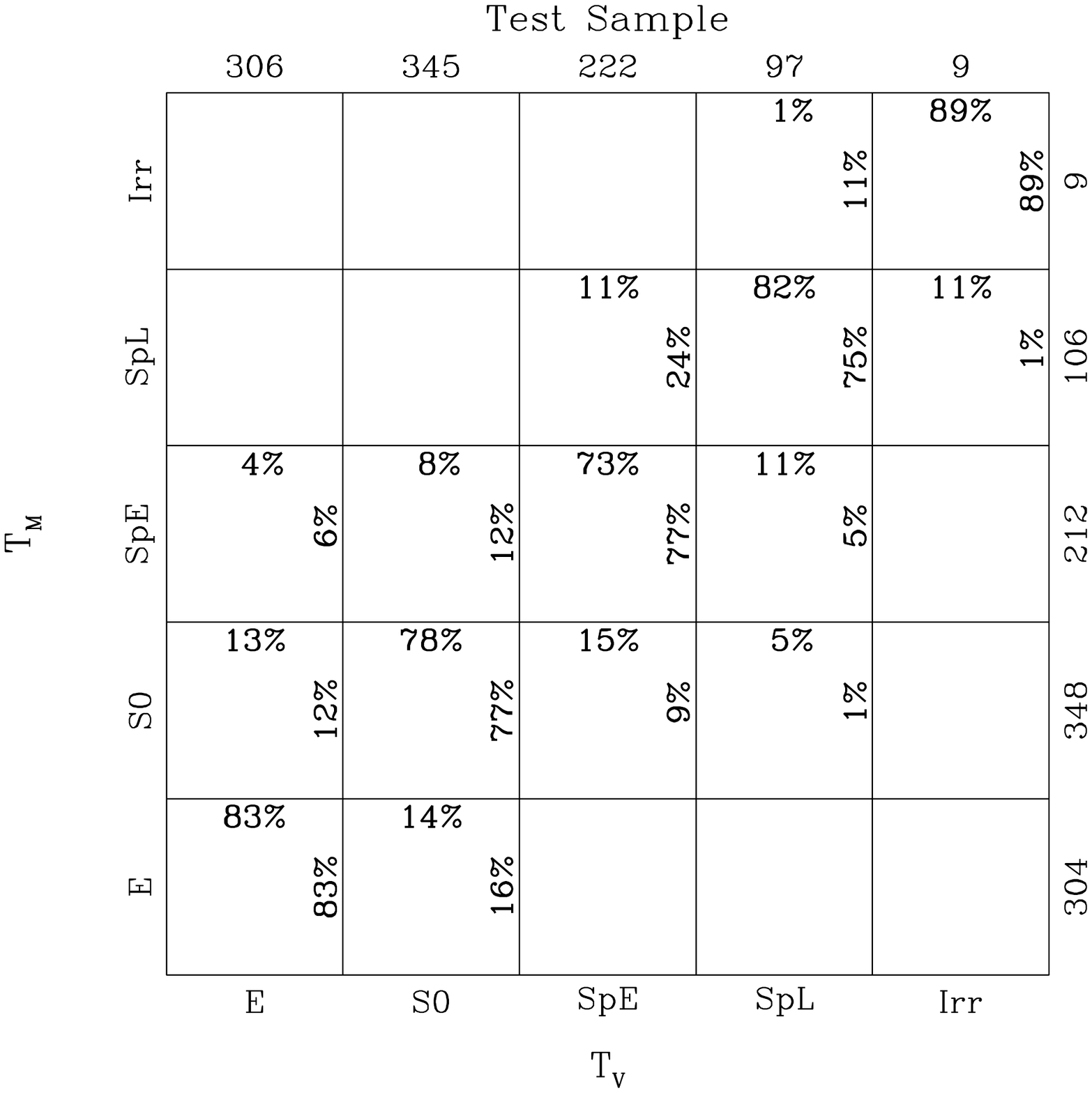}
   \vspace*{-3truecm}
     \caption{
Comparison between visual and MORPHOT 'broad' morphological
classes for the galaxies of the MORPHOT test sample. 
At the top of the 2D-bins the percentages of the visual
classes (Es, S0s, early spirals [SpE], late spirals [SpL] and
irregulars) falling in different bins of the MORPHOT
classification are reported. Similarly, the percentages of the MORPHOT
classes falling in different bins of the visual
classification are reported in the right side of the 2D-bins. Finally,
on the top (columns) and on the right (rows) of the plot, we report
the total number of galaxies in each 'broad' class of the visual and
MORPHOT estimates, respectively.}
         \label{broadcomp}
   \end{figure}

Figure~\ref{broadcomp} illustrates the
comparison between the visual and the final ($T_M$) morphological
classifications of the galaxies in the test sample. In this case
the comparison is made in 'broad' bins of
morphology, where the 'broad' classes are conventionally defined as
follows: Ellipticals (E), for $T<$-4;
Lenticulars (S0), for -4$\leq T\leq$0; Early-Spirals (SpE), for 0$<
T\leq$4; Late-Spirals (SpL), for 4$< T\leq$7; Very Late-Spirals and
Irregulars (Irr), for $T>$7.  Note that we have included the cD
galaxies in the broad class E. This is because MORPHOT tends to
classify as cDs ($T$=-6) some among the brightest and largest
ellipticals in the test sample.  Note also that we have included in
the S0 'broad' class both the galaxies classified E/S0 ($T$=-4) and
those classified S0/a ($T$=0). 

  \begin{figure*}
   \centering
   \hspace*{-2truecm}
    \includegraphics[width=16.5truecm,angle=-90]{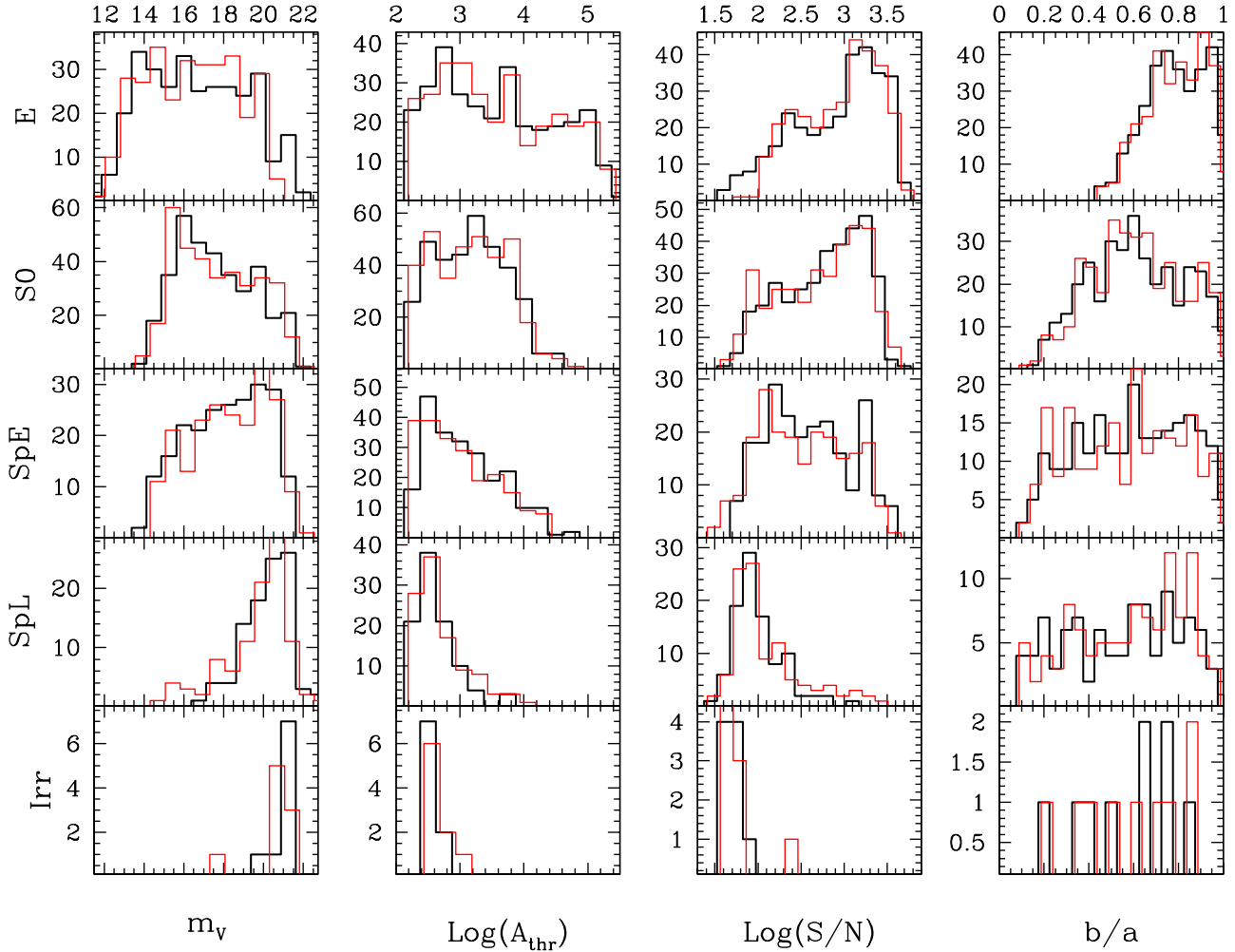}
      \caption{The distributions of apparent magnitudes (V-band),
threshold areas (pixels), signal to noise ratios and axial ratios for
different 'broad' morphological classes (Es, S0s, early spirals [SpE], 
late spirals [SpL] and irregulars) in the WINGS test galaxy sample. 
The thick line histograms refer to the visual morphological types 
($T_V$), while the thin line histograms {\bf (red in the electronic version of the paper)} illustrate the distributions 
for the MORPHOT estimates ($T_M$)}
         \label{Mres}
   \end{figure*}

Finally, Figure~\ref{Mres} illustrates in more detail the results
already shown in the previous two figures and compares, for different
'broad' morphological classes, the distributions of apparent
magnitude, threshold area, signal to noise and axial ratios obtained
from the visual (thick lines) and MORPHOT (thin lines, {\bf red in the electronic
version of the paper}) estimates of
the test galaxy sample. 

Figures~\ref{broadcomp} and \ref{Mres} show that, besides mimiching
the statistics $<\Delta T>$ and $\sigma_{\Delta T}$ of the comparison
between human classifiers (see Tables~\ref{compvis} and
~\ref{compvis1}), the automatic morphological types are able to fairly
reproduce the global morphological fractions of the visual types
(Figure~\ref{broadcomp}), as well as the fractions binned according to
several observed quantities (Figure~\ref{Mres}). {\bf In particular,
Figure~\ref{broadcomp} shows that, in spite of the wide range of
$T_M$ corresponding to each bin of $T_V$ (and viceversa), the
global fractions of visually classified E and S0 galaxies in the test sample
are almost exactly reproduced by the MORPHOT types.}

\section{Applying MORPHOT to the WINGS clusters}\label{secwings}

The bottom part of Figure~\ref{FC1} illustrates the flow-chart
relative to the application of MORPHOT to the WINGS clusters. 

Concisely: (i) for the galaxy catalog of a given WINGS cluster,
post-stamp frames of the galaxies to be classified are extracted from
the original WINGS imaging; (ii) for each galaxy, the global
quantities (R$_{80}$, $S/N$ and $\varepsilon$) and the diagnostics
$D_i$ are evaluated and the (independent) NN and ML estimators of the
morphological type are produced, each one with the proper confidence
interval; (iii) the final MORPHOT estimator is obtained by
averaging the NN and ML estimators.

According to \citet{vare09}, the WINGS optical (B,V) imaging provides
photometric and geometric description of 400140 galaxies in 77
clusters ($\sim$5200 galaxies per cluster, on average). As already
mentioned in Section~\ref{calsamp}, for about one tenth of them ({\bf 42297}
galaxies: those with threshold area greater than 200{\bf /300 pixels for images from WFC@INT/WFI@ESO}) the
surface photometry has been performed with GASPHOT \citep{pign06}. We
use the GASPHOT-WINGS catalogs as input galaxy sample to perform the
morphological analysis {\bf of WINGS galaxies} with MORPHOT.
 
\subsection{The WINGS morphological catalogs}\label{secats}

The total number of galaxies in the GASPHOT catalogs is 42,297.  We
removed from the sample those galaxies for which the Sersic index
provided by GASPHOT coincides with the boundaries of the allowed range
(0.5$-$8), which usually indicates that the fitting procedure was
unsuccessful \citep{pign06}. In this way we are left with 39651
galaxies in the fields of the WINGS clusters. For 527 galaxies (1.3\%
of this sample) MORPHOT produced unreliable results since it was not
able to compute some of the diagnostics (fuzzy/faint objects). The
remaining 39124 galaxies have been processed by MORPHOT, which
provides $T_{ML}$, $T_{NN}$ and $T_M$ estimates of the morphological
types, together with the corresponding confidence intervals $T^{min}$
and $T^{max}$.

We have very quickly checked on the WINGS imaging the MORPHOT
classification of the bright galaxies (mostly brighter than $V$=18)
and in 426 cases ($\sim$1\% of the total sample; $\sim$5\% of the
checked sample) we have modified the final classification since it was
clearly wrong. Profiting from this visual inspection procedure, we
have also manually added to the catalogs the morphological types of
799 bright galaxies (again mostly brighter than $V$=18), close to the 
borders of the frames or close to bright stars, {\bf which had been discarded 
'a priori' by the GASPHOT and MORPHOT tools.} 
After this manual intervention, the total number of WINGS
galaxies for which we provide the final morphological type estimate
($T_F$) is 39923. Among them, 2963 are visual estimates ($T_V$).  This
latter number includes the 926 galaxies of the Calibration Sample, the
979 galaxies of the Test Sample (136 of them turned out to {\bf be in common
with} the calibration sample; see Section~\ref{secperf}), the 426
galaxies whose classification has been modified after visual check (31
of them turned out to {\bf be in common with} the test sample) and the 799
galaxies manually added to the catalogs. The full MORPHOT catalog of
WINGS galaxies is available from the ``Centre de Donn\'ees
Astronomiques de Strasbourg'' (CDS) using the ViZiER Catalogue
Service. Table~\ref{morcat} show the first few records of the catalog.

\begin{table*}
\caption{Sample rows of the MORPHOT WINGS catalog}             
\label{morcat}      
\centering                          
   \hspace*{-1truecm}
\begin{tabular}{c r r r r r r r r r r r r} 
\hline\hline                 
   WINGS\_ID & Cluster & $T_{ML}$ & $T_{ML}^{min}$ & $T_{ML}^{max}$ & $T_{NN}$ & $T_{NN}^{min}$ & $T_{NN}^{max}$ & $T_M$ & $T_{M}^{min}$ & $T_{M}^{max}$ & $T_V$ & $T_F$ \\
\hline                
   WINGSJ103833.76-085623.3 &     A1069 &     3.1 &    -0.3 &     4.6 &     5.9 &     1.6 &     8.6 &     4.5 &     2.7 &     6.3 &         &    4.5  \\
   WINGSJ103834.09-085719.2 &     A1069 &     4.3 &     3.0 &     5.2 &     4.1 &     0.5 &     7.7 &     4.2 &     3.7 &     4.7 &         &    4.2  \\
   WINGSJ103834.13-085030.4 &     A1069 &    -5.0 &    -5.0 &    -4.7 &    -5.1 &    -6.0 &    -4.4 &    -5.0 &    -5.5 &    -4.5 &         &   -5.0  \\
   WINGSJ103835.85-084941.0 &     A1069 &     1.8 &    -0.1 &     3.7 &    -0.2 &    -3.5 &     3.9 &     0.8 &    -0.5 &     2.1 &         &    0.8  \\
   WINGSJ103835.89-085031.5 &     A1069 &    -1.5 &    -3.5 &     0.5 &    -3.2 &    -5.1 &    -1.4 &    -2.3 &    -3.4 &    -1.2 &         &   -2.3  \\
   WINGSJ103836.38-083614.8 &     A1069 &     3.8 &     1.6 &     4.2 &     5.0 &     0.2 &     8.9 &     4.4 &     3.6 &     5.2 &         &    4.4  \\
   WINGSJ103836.95-085300.8 &     A1069 &    -5.0 &    -5.0 &    -4.5 &    -5.9 &    -6.0 &    -5.1 &    -5.5 &    -5.5 &    -4.5 &    -5.0 &   -5.0  \\
   WINGSJ103837.15-085753.4 &     A1069 &     3.2 &     0.2 &     5.3 &     4.4 &     1.1 &     7.9 &     3.8 &     2.5 &     3.5 &     3.0 &    3.0  \\
   WINGSJ103837.48-083717.2 &     A1069 &    -1.9 &    -3.8 &    -0.0 &    -3.4 &    -4.8 &    -1.6 &    -2.6 &    -3.6 &    -1.6 &         &   -2.6  \\
   WINGSJ103837.93-084940.5 &     A1069 &    -5.0 &    -5.0 &    -4.5 &    -0.9 &    -5.2 &     3.7 &    -3.0 &    -5.0 &    -1.0 &         &   -3.0  \\
   WINGSJ103838.65-084938.3 &     A1069 &    -2.4 &    -3.5 &    -1.1 &     0.5 &    -3.2 &     4.8 &    -0.9 &    -2.4 &     0.6 &         &   -0.9  \\
   WINGSJ103839.63-084742.6 &     A1069 &     2.0 &    -2.3 &     2.7 &    -1.4 &    -3.9 &     0.9 &     0.3 &    -1.5 &     2.1 &         &    0.3  \\
   WINGSJ103840.54-085041.1 &     A1069 &    -3.0 &    -4.6 &     0.8 &    -4.0 &    -5.9 &    -1.4 &    -3.5 &    -4.3 &    -2.7 &         &   -3.5  \\
   WINGSJ103840.85-085046.5 &     A1069 &    -4.9 &    -3.8 &    -2.3 &    -5.3 &    -6.0 &    -4.5 &    -5.1 &    -5.6 &    -4.6 &         &   -5.1  \\
   WINGSJ103841.43-085521.6 &     A1069 &    -3.7 &     1.5 &     3.3 &     6.0 &     0.9 &     9.4 &     1.1 &    -3.8 &     6.0 &         &    1.1  \\
   WINGSJ103841.63-085528.6 &     A1069 &    -4.8 &    -3.5 &    -2.7 &    -3.5 &    -5.5 &    -1.4 &    -4.2 &    -4.9 &    -3.5 &         &   -4.2  \\
   WINGSJ103842.12-083557.8 &     A1069 &    -5.0 &    -5.0 &    -4.1 &     0.5 &    -4.5 &     6.0 &    -2.3 &    -5.0 &     0.4 &         &   -2.3  \\
   WINGSJ103843.03-085602.8 &     A1069 &    -5.0 &    -5.0 &    -4.3 &    -4.9 &    -5.8 &    -3.8 &    -5.0 &    -5.5 &    -4.5 &         &   -5.0  \\
   WINGSJ103844.19-085609.3 &     A1069 &     0.5 &    -2.1 &     0.7 &    -1.9 &    -3.8 &     0.4 &    -0.7 &    -2.0 &     0.6 &         &   -0.7  \\
   WINGSJ103844.58-084601.1 &     A1069 &     4.5 &     3.4 &     5.4 &     5.3 &     1.2 &     8.4 &     4.9 &     4.3 &     5.5 &         &    4.9  \\
   WINGSJ103845.53-085341.2 &     A1069 &     3.1 &     1.4 &     4.0 &    -3.1 &    -5.3 &    -0.5 &     0.0 &    -3.1 &     3.1 &         &    0.0  \\
   WINGSJ103845.73-084021.9 &     A1069 &     0.3 &    -1.6 &     0.6 &    -3.8 &    -6.0 &     1.9 &    -1.7 &    -3.8 &     0.4 &         &   -1.7  \\
   WINGSJ103846.39-084530.6 &     A1069 &    -3.9 &    -4.8 &    -0.9 &    -2.3 &    -4.8 &     0.7 &    -3.1 &    -4.5 &    -3.5 &    -4.0 &   -4.0  \\
   WINGSJ103846.47-084226.0 &     A1069 &    -0.9 &    -2.6 &     0.8 &     0.7 &    -1.7 &     2.9 &    -0.1 &    -1.2 &     1.0 &         &   -0.1  \\
   WINGSJ103847.59-084631.3 &     A1069 &    -1.4 &    -2.1 &    -0.2 &    -5.5 &    -6.0 &    -4.3 &    -3.5 &    -5.6 &    -1.4 &         &   -3.5  \\
   WINGSJ103848.08-085044.8 &     A1069 &    -5.0 &    -5.0 &    -4.5 &    -3.9 &    -4.6 &    -2.9 &    -4.4 &    -5.0 &    -3.8 &         &   -4.4  \\
   WINGSJ103848.30-084259.9 &     A1069 &    -0.1 &    -2.2 &     1.1 &     0.3 &    -1.6 &     2.3 &     0.1 &    -0.5 &     0.7 &         &    0.1  \\
   WINGSJ103850.17-085336.6 &     A1069 &    -2.7 &    -0.5 &     2.2 &     2.3 &    -1.5 &     6.3 &    -0.2 &    -2.7 &     2.3 &         &   -0.2  \\
   WINGSJ103850.35-084804.5 &     A1069 &    -2.8 &    -3.3 &    -0.8 &    -0.5 &    -4.0 &     3.8 &    -1.7 &    -3.0 &    -0.4 &         &   -1.7  \\
\hline                                   
\end{tabular}
\end{table*}

\subsection{External comparisons}

In order to provide an external check of the goodness of our automated
classifications, we have searched the literature for visually
classified galaxy samples having objects in common with our
WINGS-MORPHOT sample. We only found three possible data samples that
could be usable for our purpose, all of them concerning the SDSS
galaxies: \citet{fuku07}, \citet{nair10} and \citet[][Galaxy
Zoo]{lint11}.  By cross-matching these samples with our catalog, we
found that the objects in common are 18, 79 and 2110,
respectively. However, the potentially most sizeable comparison sample
(Galaxy Zoo) turns out to be practically useless, since it just
provides the binary information Elliptical/Spiral (no S0
classification). Moreover, while the morphological resolution of the
classification system adopted by \citet{nair10} is comparable (not
equal) to the MORPHOT resolution, the \citet{fuku07} system only
enables us to compare the 'broad' morphological classes.

   \begin{figure}
   \hspace*{-1truecm}
   \vspace*{-1truecm}
  \centering
   \includegraphics[width=8truecm,angle=-90]{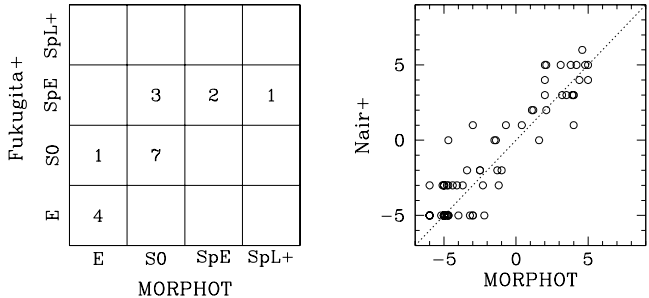}
      \caption{Comparison between MORPHOT and literature morphological types
(see the comments in the text).
{\it Left panel}: number of galaxies in the different bins of 'broad' morphological
class for the sample of 18 WINGS-MORPHOT galaxies in common with \citet{fuku07}. 
{\it Right panel}: the MORPHOT vs. \citet{nair10} morphological types for the 79
common galaxies.}
        \label{EXTCOMP}
   \end{figure}

Figure~\ref{EXTCOMP} illustrates the comparison of the MORPHOT
morphological types with the \citet{fuku07} and \citet{nair10} 
classifications (left and right panel, respectively). In both cases the MORPHOT 
results tend to be slightly shifted towards earlier types with
respect to the visual estimates. However, in spite of the
small number of cross-matched galaxies and of the different
classification systems, the agreement between the MORPHOT 
automated morphological types and the visual estimates from the
literature looks satisfactory. In particular, the average value
and the scatter of the difference ($T_M$-$T_{Nair+}$) turn out
to be -0.4 and 1.4, respectively, to be compared with the 
corresponding values given in Tables~\ref{compvis} and \ref{compvis1}.

\subsection{Morphological properties of the WINGS clusters}\label{secprop}

This section outlines the main statistical properties of the
galaxy morphology in the WINGS clusters. More detailed and 
exhaustive analyses about this topic will be presented in a few
forthcoming papers. Here we just illustrate some general 
morphological trend emerging from WINGS-MORPHOT catalogs. 

\begin{figure}
   \hspace*{-0.5truecm}
   \centering
   \includegraphics[width=9.5truecm]{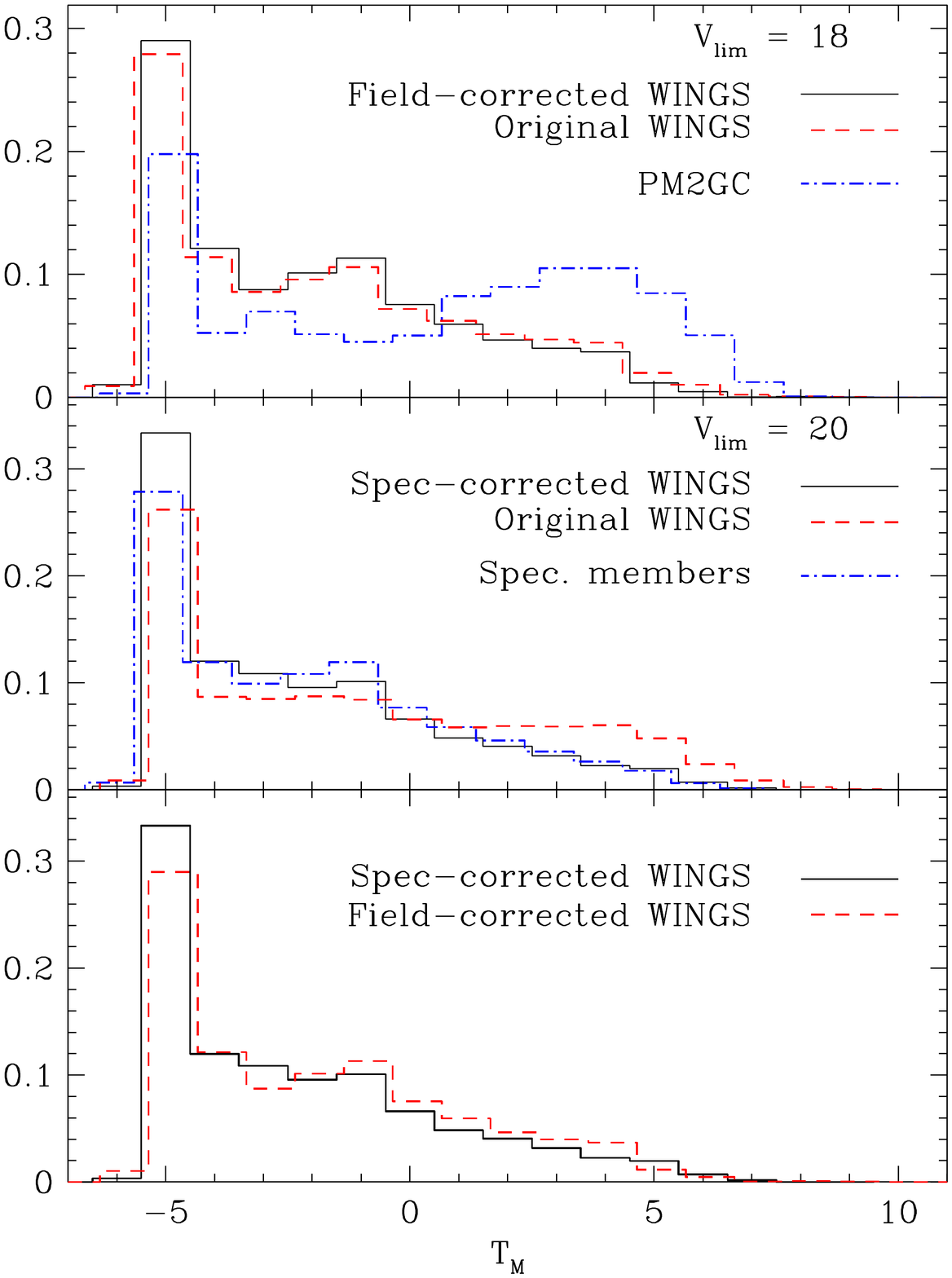}
      \caption{Distribution of MORPHOT types in the WINGS clusters.
{\it Upper panel}: The distributions for the uncorrected WINGS sample, 
for the general field sample \citep[from PM2GC;][]{calv11} and for the
field-corrected WINGS sample up to the apparent magnitude V=18
are illustrated by dashed, dot-dashed and full line histograms, respectively.
{\it Middle panel}: The distributions for the uncorrected WINGS sample, 
for the spectroscopic WINGS members and for the
WINGS sample corrected with the spectroscopic copleteness and membership
functions derived from \citet{cava09} are illustrated by dashed, dot-dashed 
and full line histograms, respectively. In this case a deeper magnitude
limit (V=20) has been used (see text).
{\it Bottom panel}: Comparison between the field-corrected and the
spectroscopy-corrected distribution of the MORPHOT types in the WINGS
clusters. In spite of the different magnitude limits adopted for the
statistical corrections, the two distributions turn out to be quite consistent
each one another. The same happens if we adopt for the spectroscopic
membership correction the same magnitude limit we used for the
statistical filed (PM2GC) correction ($V_{lim}$=18).}
        \label{TDISTR}
   \end{figure}

Figure~\ref{TDISTR} illustrates the distribution of the MORPHOT types
($T_M$) in the fields of the WINGS clusters. The contribution of the
general field to the different morphological types has been estimated
in two independent ways. First, we have counted galaxies in the
general field of the Padova Millennium Galaxy and Group Catalogue
(PM2GC; \citealp{calv11}), for which we have
obtained MORPHOT classifications. Second, we have estimated the
fraction of cluster members in the WINGS fields using the
spectroscopic completeness and membership functions derived for the
WINGS survey by \citet{cava09}. In the first case (upper panel of
Figure~\ref{TDISTR}) we can confidently assume the PM2GC survey to be
nearly complete down to $V_{lim}$=18, while the selection of the WINGS
spectroscopic sample has been extended down to $V\sim$20 (central
panel of the figure).  In this case, the completeness function is
mainly determined by time allocation and fiber crowding
problems. Moreover, the spectroscopic WINGS survey only includes a
subsample of the original WINGS cluster sample (48 over 77). In spite
of these differences, the distributions of the MORPHOT types in the
WINGS clusters, obtained applying to the WINGS-MORPHOT catalogs the
two (independent) statistical corrections for membership, turn out to
be remarkably consistent (see the bottom panel in the same figure).
The consistency is confirmed even if we use $V_{lim}$=18 also for the
membership correction based on the spectroscopic completeness.

\begin{figure}
   \hspace*{-0.5truecm}
   \vspace*{-1truecm}
  \centering
   \includegraphics[width=7.5truecm,angle=-90]{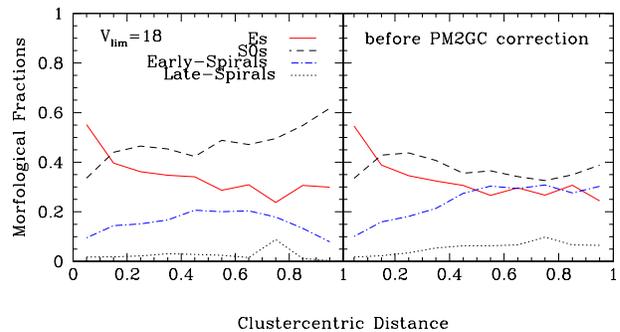}
      \caption{Morphological fractions as a function of the 
clustercentric distance (in $R_{200}$ units) for the WINGS galaxies before 
and after correction for field contamination. The general field morphological 
fractions in different bins of apparent V-band magnitude have been derived 
from the PM2GC sample \citep{calv11} using the magnitude limit V=18.}
        \label{CCDISTR}
   \end{figure}

Adopting the 'broad' morphological classes conventionally defined in
Section~\ref{secperf}, we find that ellipticals, S0s and spiral
galaxies constitute $\sim$33\%, 44\% and 23\% of the whole galaxy
population in the WINGS clusters.
 It is worth to note that these morphological fractions are slightly different from
those found in \citet{pogg09}. The discrepancy
mostly concerns the E/S0 ratio and is due to the combined effects of
two factors: (i) the limit of 0.6$\times R_{200}$ adopted for the
clustercentric distance in \citet{pogg09}; (ii) the behaviour of the
E/S0 ratio as a function of the clustercentric distance.
Figure~\ref{CCDISTR} illustrates the point, showing the morphological
fractions as a function of the clustercentric distance before and
after correction for field contamination (right and left panels,
respectively). Note the prevalence of Es in the inner cluster regions,
which is responsible for the different E/S0 ratio found in \citet{pogg09}.
Note also that the field correction does not influence this ratio up to
$\sim$0.5$\times R_{200}$, mostly operating on the
S0/Early-Spirals fraction in the external part of the clusters.

\begin{figure}
   \hspace*{-0.5truecm}
   \centering
   \includegraphics[width=8truecm]{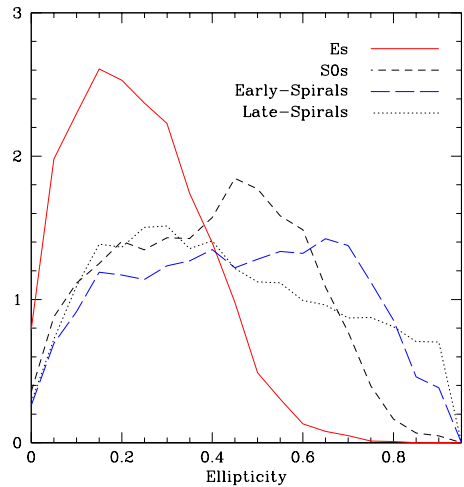}
   \vspace*{-2truecm}
      \caption{Ellipticity distributions of WINGS galaxies for different 'broad' 
morphological classes. The Irregular galaxies have been included in the Late-Spirals
(see comments in the text). No correction for field contamination has
been applied.}
        \label{EDISTR}
   \end{figure}

Figure~\ref{EDISTR} shows the distributions of the projected
ellipticities in the WINGS-MORPHOT catalogs for different 'broad'
morphological classes. In this case we do not try to correct for
cluster membership, since we just aim to check the plausibility of the
distributions, also in comparison with the literature. To this
concern, the ellipticity distribution of elliptical galaxies in
Figure~\ref{EDISTR} turns out to be in perfect agreement with
\citet[][see also \citealp{vulc11}]{fasa91}, while for S0s and
Early-Spirals the peaks of the distributions are slightly shifted
towards lower values of the ellipticities with respect to the
corresponding distributions in \citet{fasa93}. However, this is not
surprising, since our flattenings come from global, single component
(Sersic) fitting of the galaxy image (GASPHOT), while the axial ratios
in \citet{fasa93} refer to the outer isophotes (essentially the disk
components).  The peculiar ellipticity distribution of Late-Spirals is
due to the inclusion in this 'broad' class of the irregular objects,
which could be intrinsically less flattened than disk galaxies.

\begin{figure}
   \hspace*{-0.5truecm}
   \centering
   \includegraphics[width=8truecm]{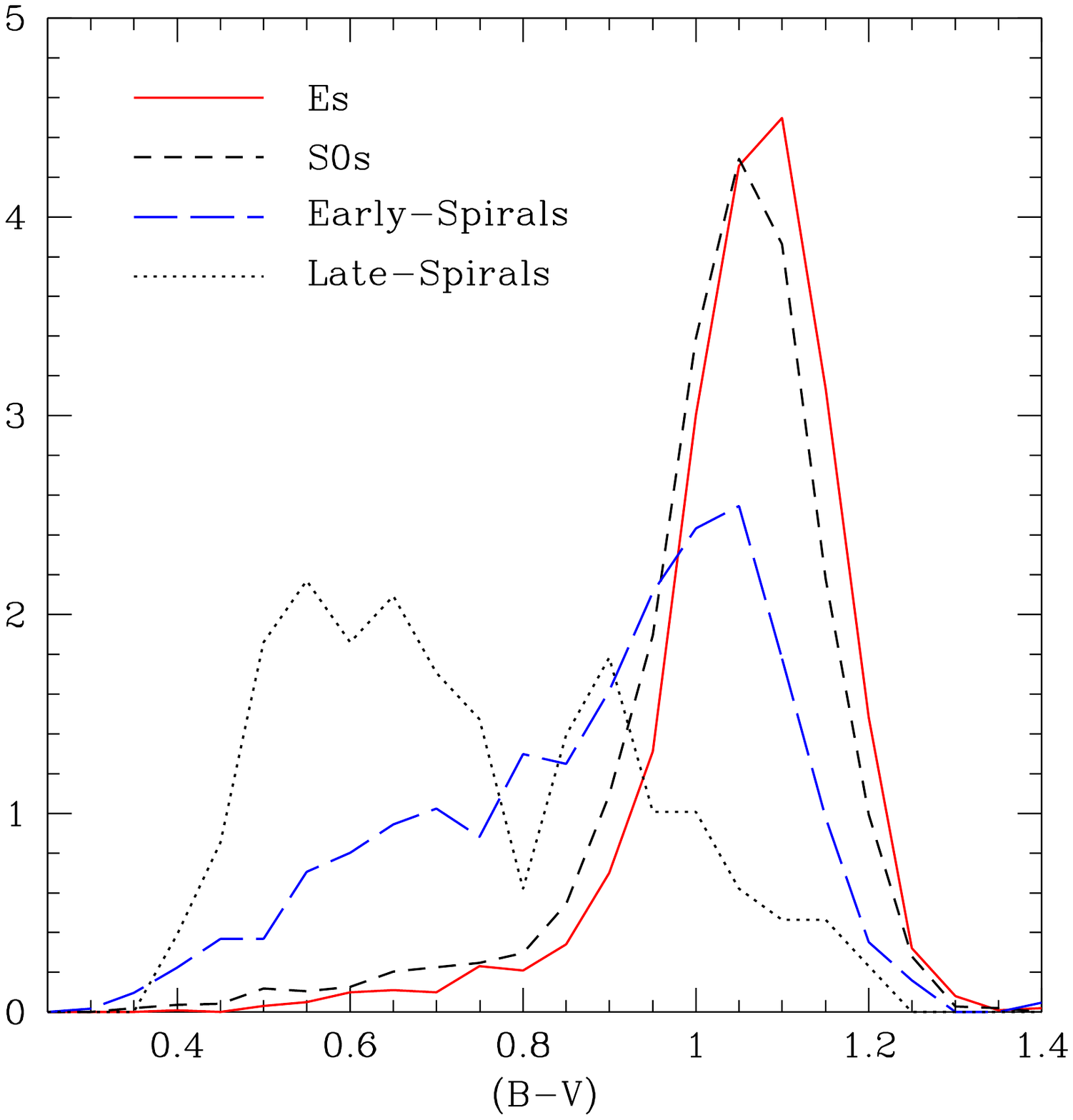}
   \vspace*{-2truecm}
      \caption{(B-V) color distributions of spectroscopically confirmed
WINGS cluster members for different 'broad' morphological
classes. The Irregular galaxies have been included in the Late-Spirals
(see comments in the text).}
        \label{CDISTR}
   \end{figure}

Figure~\ref{CDISTR} show the distributions of the (B-V) color for the
spectroscopically confirmed members in the WINGS-MORPHOT catalogs and
for the different 'broad' morphological classes. Note the remarkable
similarity and the small, but statistically significant shift between
the distributions of E and S0 galaxies. Note also the bi-modal color
distribution of the Late-Spirals(+Irr) galaxies.

\begin{figure}
   \hspace*{-0.5truecm}
   \centering
   \includegraphics[width=8truecm]{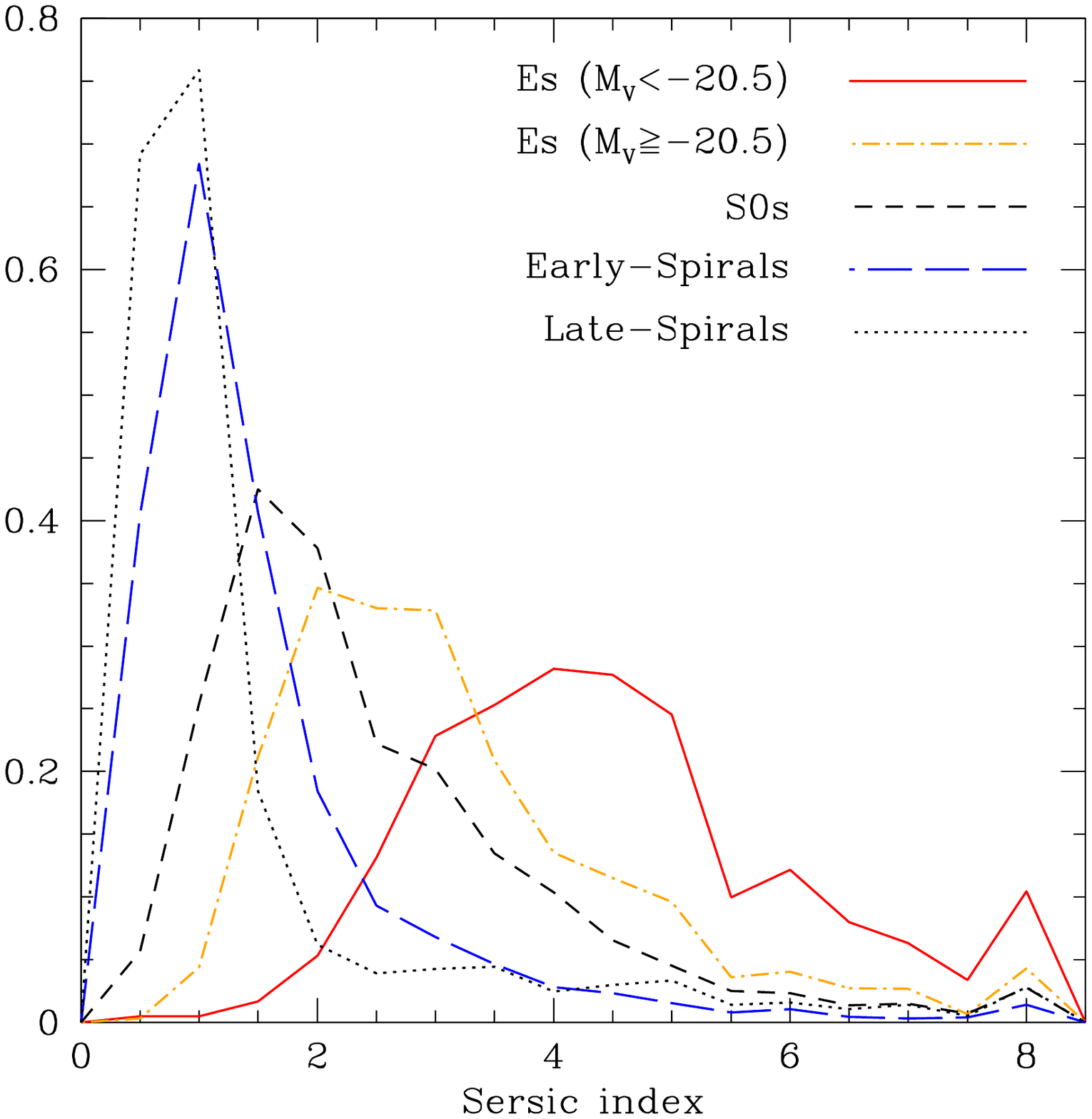}
   \vspace*{-2truecm}
      \caption{
        Sersic index distributions of WINGS galaxies for different
        'broad' morphological classes. Elliptical galaxies have been
        divided in bright ($M_V<$-20.5) and faint ($M_V\ge$-20.5) Es,
        while the Irregular galaxies have been included in the
        Late-Spirals (see comments in the text). No correction for
        field contamination has been applied.}
        \label{NDISTR}
   \end{figure}

Finally, in Figure~\ref{NDISTR} we present the distribution of the
Sersic index $n$ for the different 'broad' morphological classes.
Again, in this case we do not try to correct for cluster membership.
The Sersic indices come from our WINGS-GASPHOT catalogs.  As already
mentioned in Section~\ref{secintro}, even though a correlation between
these {\it n} and the morphological type exists, it is weak and it
shows a high degree of degenaracy, especially for early-type galaxies
(see the distributions for bright and faint Es).

\section{Summary}\label{secsum}

In this paper we have presented the morphological classification of
$\sim$40000 galaxies in the fields of 76 nearby clusters from the
WINGS optical (V-band) survey. The morphological types have been
estimated automatically, using the purposely devised tool MORPHOT,
whose description takes up a substantial part of the paper. It
combines a large set (21) of diagnostics, easily computable from the
digital cutouts of galaxies, producing two different estimates of the
morphological type based on: (i) a semi-analytical Maximum Likelihood
technique; (ii) a Neural Network machine. The final, averaged
estimator has been tested over a sample of $\sim$1000 visually
classified WINGS galaxies, proving to be almost as effective as the
'eyeball' estimates themselves. In particular, at variance with most
existing tools for automatic morphological classification of galaxies,
MORPHOT has been shown to be able to distinguish between ellipticals
and S0 galaxies with unprecedented accuracy.  Even though its basic
methodology is robust for any set of digital images of similar spatial
resolution and dynamic range, MORPHOT is presently calibrated and
fine-tuned to provide reliable morphologies of WINGS galaxies
alone. Adjustments of the calibration are required (and are actually
in progress) to make the tool more generally usable.  The
WINGS-MORPHOT catalog has been exploited here just to illustrate the
distributions of some relevant photometric and structural properties
of galaxies in the WINGS clusters. In a few forthcoming papers of the
WINGS series, we plan to perform more detailed statistical analyses
involving the morphology of cluster galaxies. In particular, besides
the classical morphology--density and morphology--clustercentric
distance relations, we will exploit the WINGS spectroscopic
information \citep{cava09,frit07,frit11,hans11} to study how galaxy
morphology correlates with star formation rate and history at
different clustercentric distances.

\section*{Acknowledgments}

{\bf We thank the referee Roberto Abraham for the patience and the carefulness in reading
our heavy (boring!) paper and for the few, but useful suggestions.}

\appendix

\section{The current set of diagnostics: definitions}\label{appdia}

Here we present in some detail the definition and the meaning of the
20 diagnostics $D_i$ (i=1,...,20) we devised up to now. The first nine of the following
diagnostics are actually already present in the literature, although
sometimes in slightly different forms. We will refer to the original
papers for details about their definitions. Instead, the remaining
eleven diagnostics are presented here for the first time. Hereafter,
in the definition of diagnostics, we use just the pixels above the 
threshold value (2$\sigma_{bkg}$) and far from the galaxy center 
more than the image FWHM.

\subsection{Diagnostics already present in the literature}\label{olddia}

{\bf $D_1$: Sersic index of the luminosity profile}.

\noindent
Given the FWHM, this diagnostic is evaluated on the image $F_S$
according to the prescriptions given in \citet[][; Section~4
therein]{truj01}, making use of the previously extracted elliptical
aperture intensity profile of the galaxy (see Section~\ref{preproc});

\smallskip\noindent
{\bf $D_2$: Luminosity-ranked Concentration index}.

\noindent
Again from the image $F_S$ and from the elliptical aperture intensity
profile, this diagnostic is evaluated as the fraction of the total
intensity coming from the 30\% brightest pixels;

\smallskip\noindent
{\bf $D_3$: Distance-ranked Concentration index}.

\noindent
Similar to the previous one, but defined as the fraction of the total
intensity coming from the 30\% pixels closest
to the galaxy center (in units of elliptical distances). Note that
more elaborated versions of the Concentration indices can be found in
\citet{grah01}, \citet{cons03} and \citet{yama05};

\smallskip\noindent
{\bf $D_4$: Luminosity-ranked Gini Coefficient}.

\noindent
Following \citet{abra03} and \citet{lotz04}, inside the square whose
sides coordinates are the fraction of galaxy pixels and the fraction
of the total counts (square area $\equiv$ 1), we define this
diagnostic as the area between the diagonal of the square and the
galaxian Lorentz curve (i.e.: the rank-ordered cumulative distribution
function of the pixel counts).  For this diagnostic and for the
following one we use the image $F_S$;

\smallskip\noindent
{\bf $D_5$: Distance-ranked Gini Coefficient}.

\noindent
Similar to the previous one, but in this case the pixels are ranked in
ascending order of elliptical distance from the galaxy center;

\smallskip\noindent
{\bf $D_6$: Second-order moment of light}.

\noindent
Following \citet{lotz04}, we define this diagnostic as the
second-order moment of the brightest 20\% pixels of the image $F_S$,
normalized to the same moment computed over the whole galaxy area;

\smallskip\noindent
{\bf $D_7$: Asymmetry}.

\noindent
For this diagnostic we use the image $F_C$ and we adopt the definition
given in \citet[][; normalized square counts of the difference between
the original and the 180$^\circ$ rotated image]{cons97}, with the
improvements suggested in \citet[][; see their Section~3.4]{cons00}
about the preliminar image processing (careful centering) and the
handling of the uncorrelated noise;

\smallskip\noindent
{\bf $D_8$} and {\bf $D_9$: Clumpiness}.

\noindent
The diagnostic $D_8$ is defined according to \citet[][see
equation~2 therein]{cons03}, including in the sum just the pixels
with counts above the threshold (2$\sigma_{bkg}$) and far from the
galaxy center more than the image FWHM. The second clumpiness
diagnostic ($D_9$) is similar to the previous one. However, in
order to further enhance high-frequency features, in this case the
model image $F_M$ defined in Section~\ref{preproc} is subtracted from
the galaxy image $F_C$, instead of the gauss-smoothed version of the
image itself;

\subsection{New Diagnostics}\label{newdia}

The following morphological diagnostics are presented here for the
first time. They all are computed on the frame $F_C$, again using just the
pixels above the threshold (2$\sigma_{bkg}$) and far from the galaxy
center more than the image FWHM.  We recall (see
Section~\ref{calsamp}) that $F_C(i,j)$ is a square matrix, whose
size (N) must be an odd number.  

\smallskip\noindent
{\bf $D_{10}$} and {\bf $D_{11}$: Diskyness}.

\begin{figure}
   \vspace*{-1.5truecm}
   \centering
   \hspace*{-1truecm}
   \includegraphics[width=10truecm]{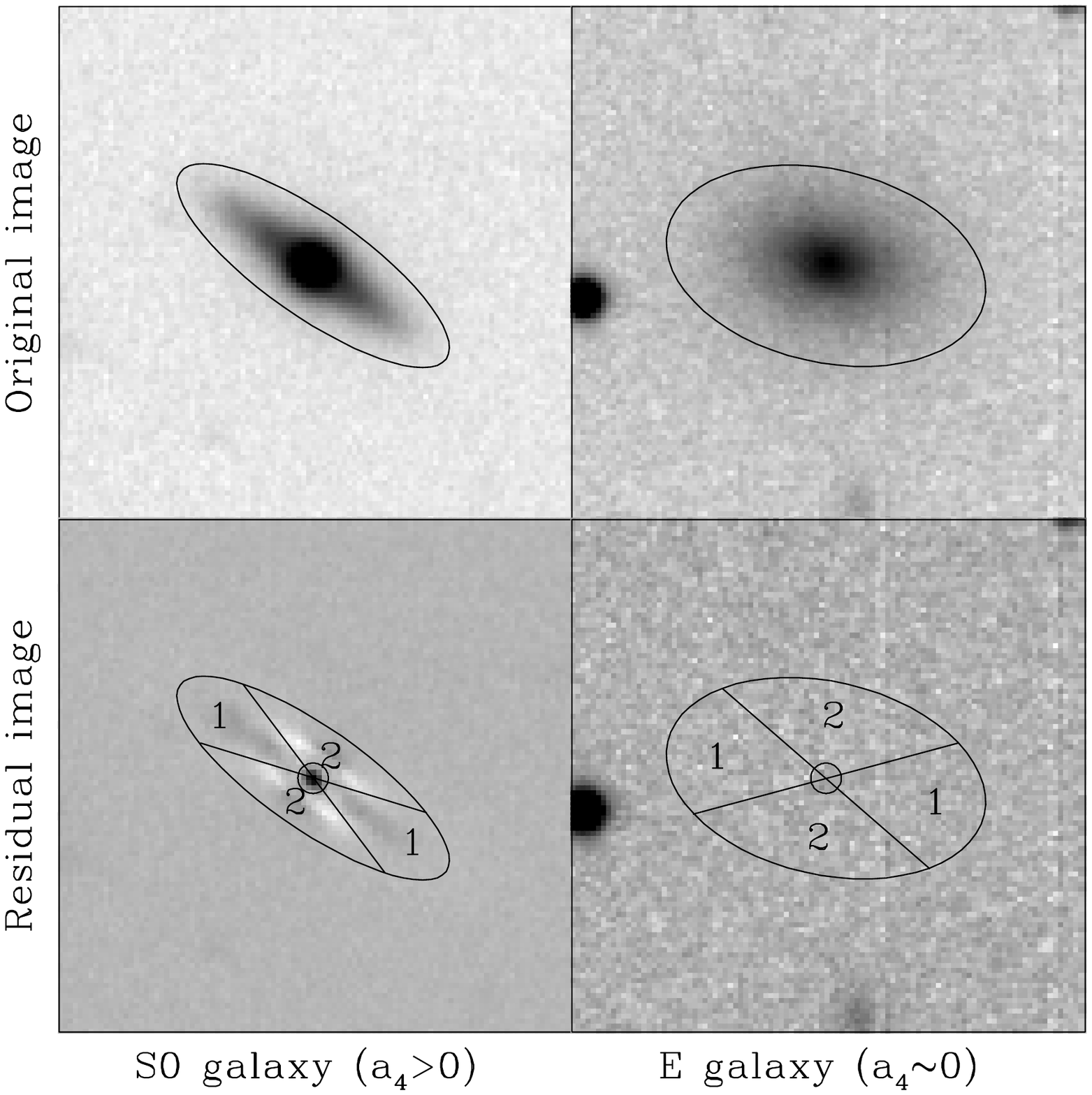}
   \vspace*{-3truecm}
      \caption{Illustration of the diagnostic $D_{10}$
(see text for a detailed explanation).}
         \label{disk1}
   \end{figure}

\noindent
The difference between the galaxy image $F_C$ and the model image
$F_M$ (residual image; hereafter $F_R$) is used to devise two new
diagnostics related to the shape of the galaxy isophotes
\citep[$a_4>0\equiv$disky, $a_4<0\equiv$boxy; ][]{bend87}. The first diagnostic is defined as:
$$D_{10}=(<F_{R1}>-<F_{R2}>)/<|F_R|>$$
where $<F_{R1}>$ and $<F_{R2}>$ are the average counts of $F_R$ in
the equal-area sectors of the above defined model-ellipse marked
respectively with {\bf 1} and {\bf 2} in the bottom panel of
Figure~\ref{disk1}. In the formula, the quantity $<|F_R|>$ is a
normalization factor representing the average value of the (absolute)
counts of $F_R$ over the whole model-ellipse (apart from the inner
circle of radius=FWHM).  Clearly, $D_{10}$ tend to be positive in galaxies
with disk-shaped isophotes, since in this case the residuals in the
two sectors marked {\bf 1} tend to be greater (darker in the figure) than
those in sectors marked {\bf 2}.

  \begin{figure}
   \vspace*{-3truecm}
   \centering
   \hspace*{-1truecm}
  \includegraphics[width=10truecm]{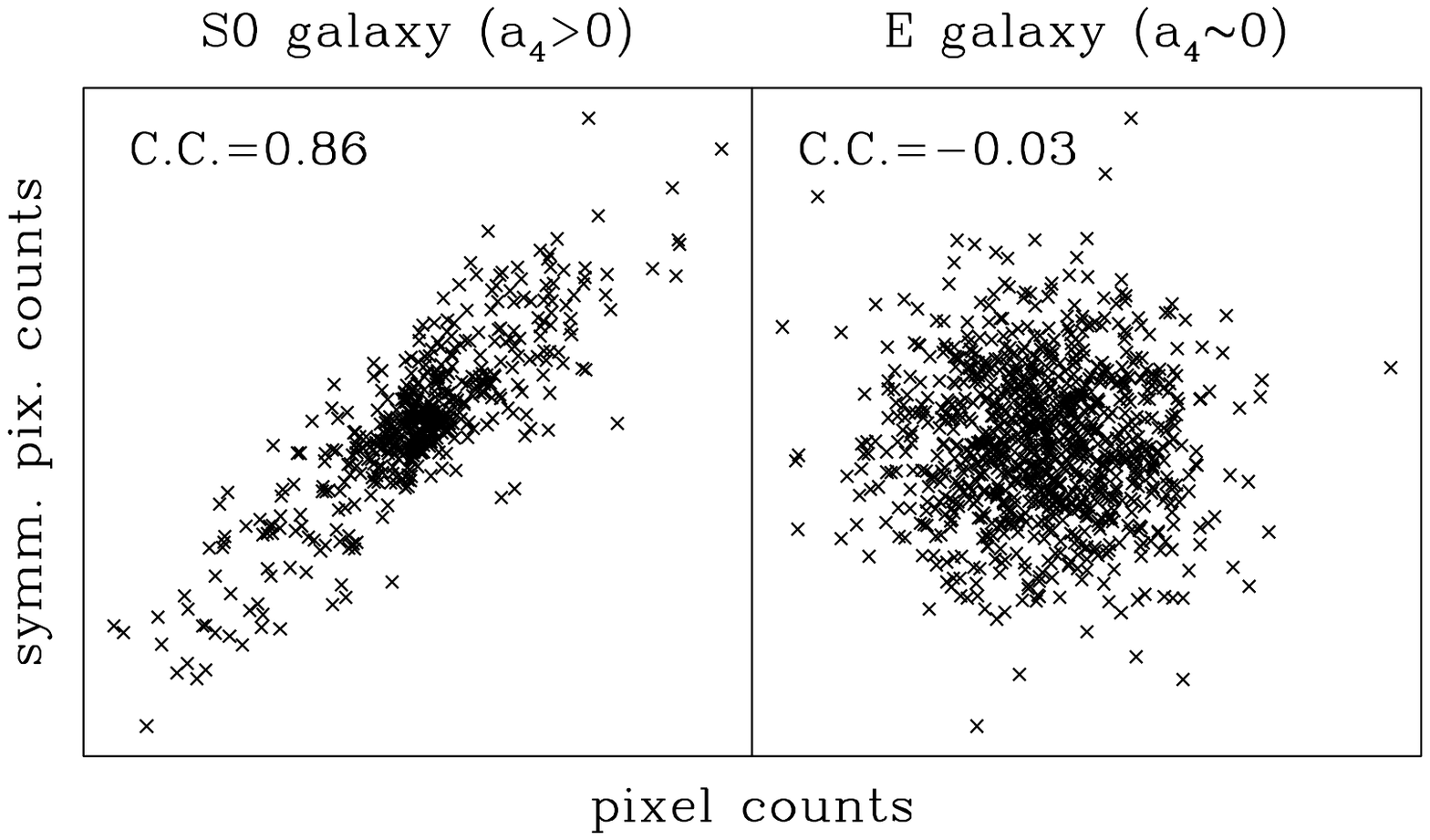}
   \vspace*{-5truecm}
      \caption{Illustration of the diagnostic $D_{11}$
(see text for a detailed explanation).}
         \label{disk2}
   \end{figure}

\noindent
The second diagnostic ($D_{11}$) is defined as the correlation
coefficient (C.C.) between the counts of the pixels within
whatever half-part of the model-ellipse (for instance the bottom-half:
Y$_{pix}\le$Y$_{cen}$) and those of the corresponding pixels symmetric
with respect to the galaxy center. In Figure~\ref{disk2} these two
quantities are plotted against each other for the same galaxies used
in Figure~\ref{disk1}, i.e. a S0 galaxy with disky isophotes (left
panel) and an elliptical galaxy (right panel). It is evident that high
values of C.C. correspond to strongly disk-shaped objects, while for 
regular (non disky) ellipticals the C.C. values are close to zero;

\smallskip\noindent
{\bf $D_{12}$: Bandwidth of Power Spectrum}.

\noindent
Roughly speaking, late--type galaxies are dominated by structural
features (spiral arms, clumps, tails, blobs, etc..) whose size is
(much) lower than galaxy size, while elliptical and (in general)
early--type galaxies are typically dominated by a single, regular
structure, whose size is comparable with that of the galaxies
themselves. Having this in mind, we have devised a new morphological
diagnostic defined as the ratio between some characteristic inverse
frequency of the 2D power spectrum of the galaxy image (i.e. the
typical size of features) and the equivalent threshold radius of the
galaxy ($\sqrt{A_{thr}/\pi}$). Operatively, we estimate the
characteristic size of galaxy features by processing the $F_C$ image
with the {\tt powerspec} tool (option: {\tt center=yes}) included in
the {\tt IRAF stsdas-Fourier} package (FFT), and computing on the {\tt
  powerspec} image the equivalent radius of the area where the power
exceeds half of the maximum FFT power.

\smallskip
The next four morphological diagnostics concern the statistical
behaviour of very local pixel properties (image texture; a similar 
approach can be found in \citealp{moor06}) over the
whole galaxy body. In particular, they consider the texture unities
($TUs$ hereafter) provided by all the 3$\times$3 pixel squares
centered on each pixel of the frame $F_C(i,j)$ for $i=j=$2,...,(N-1).
In order to make easier the formalism related to these diagnostics, it
is convinient to introduce the following definitions relative to each
$TU_{ij}$ (see Figure~\ref{pixsqr}):

  \begin{figure}
   \vspace*{-1.5truecm}
   \centering
   \hspace*{-1truecm}
   \includegraphics[width=8truecm]{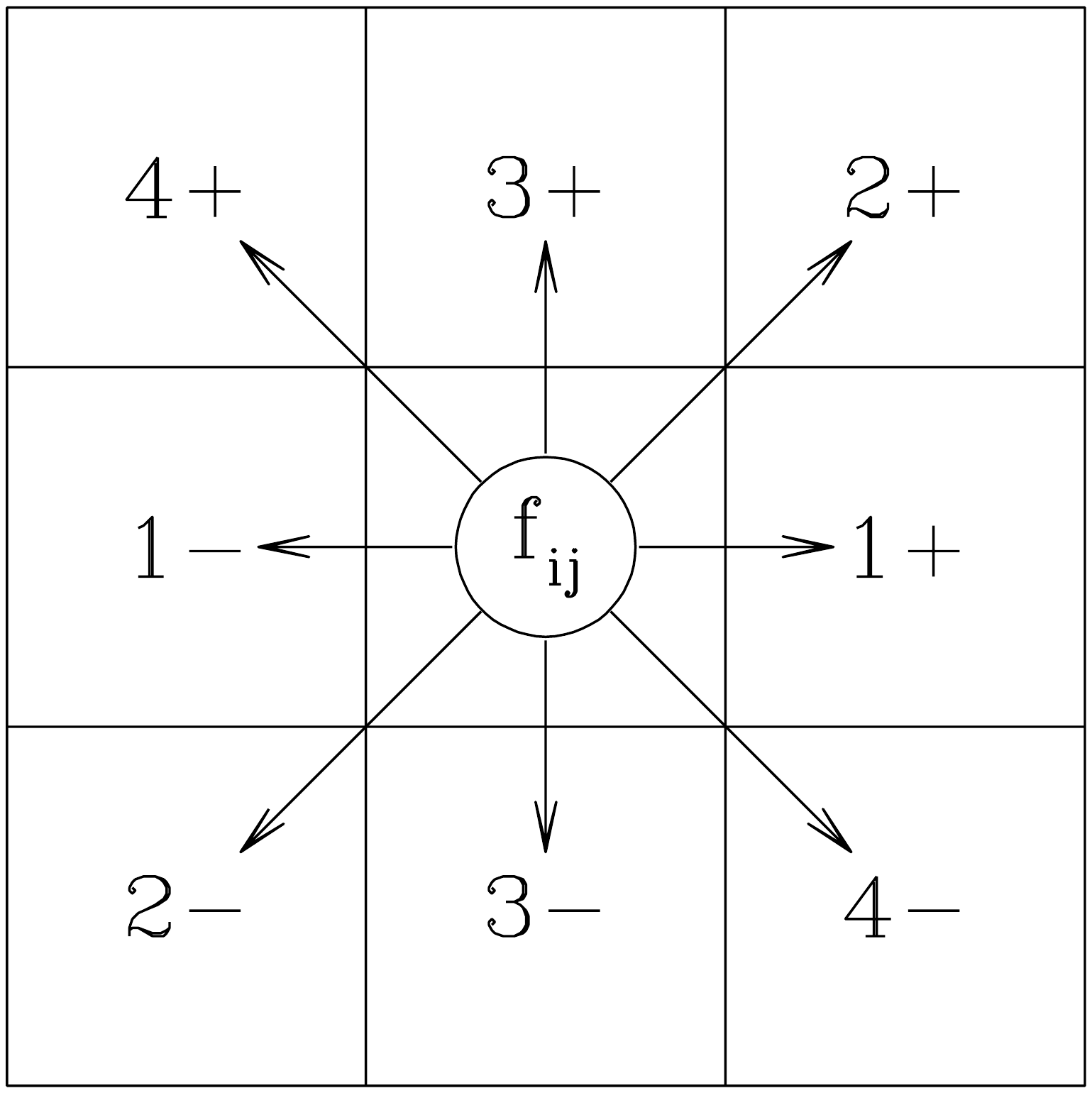}
   \vspace*{-3truecm}
      \caption{Scheme of a Texture Unit (TU$_{ij}$), illustrating the meaning
of its upper indices (see the formulation in the text).}
         \label{pixsqr}
   \end{figure}

\medskip
$f_{ij}=F_C(i,j)$; 

$f_{ij}^{1-}=F_C(i-1,j)$; $f_{ij}^{1+}=F_C(i+1,j)$; 

$f_{ij}^{2-}=F_C(i-1,j-1)$; $f_{ij}^{2+}=F_C(i+1,j+1)$; 

$f_{ij}^{3-}=F_C(i,j-1)$; $f_{ij}^{3+}=F_C(i,j+1)$; 

$f_{ij}^{4-}=F_C(i+1,j-1)$; $f_{ij}^{4+}=F_C(i-1,j+1)$; 

\smallskip
$\cos\alpha_{ij1}=\vert i-i_c\vert / d_{ij}$; $\cos\alpha_{ij2}=\vert i+j-i_c-j_c\vert / d_{ij}\sqrt{2}$;

$\cos\alpha_{ij3}=\vert j-j_c\vert / d_{ij}$; $\cos\alpha_{ij4}=\vert i-j\vert / d_{ij}\sqrt{2}$,

\medskip\noindent
where $i_c\equiv j_c\equiv$(N/2+0.5) are the coordinates of the
center, $d_{ij}=\sqrt{(i-i_c)^2+(j-j_c)^2}$ is the distance from the
center of the pixel ($i,j$) and $\alpha_{ijk}$ ($k$=1,..,4) is the angle
between the direction $(k-)\leftrightarrow(k+)$ in the $TU_{ij}$
of Figure~\ref{pixsqr} and the line connecting ($i,j$) with the
galaxy center. 

\smallskip\noindent
{\bf $D_{13}$: Average Concaveness}.

\noindent
Again roughly speaking, the earlier the morphological type, the lower
the unevennes of the intensity surface of the galaxy. Moreover, while in
early--type galaxies the intensity gradient increases regularly toward 
the center over almost the whole galaxy body, for
late--type objects, due to the presence of relatively small structural
features (spiral arms, clumps, tails, blobs, etc..), such regular
behaviour is limited to the very inner part of the galaxy (bulge).
That being stated and given the above definitions, this new diagnostic 
is expressed by the formula:
$$D_{13}=0.5+{\displaystyle\sum_{i,j=2}^{N-1}{f_{ij}\times sgn\big{\{}\sum_{k=1}^4{\cos\alpha_{ijk}(f_{ij}^{k-}+f_{ij}^{k+}-2f_{ij})\big{\}}}}\over\displaystyle{2\times\sum_{i,j=2}^{N-1}{f_{ij}}}}$$
which in fact provides the fraction of the total galaxy luminosity
coming from pixels ($i,j$) for which is positive the local
concaveness, computed in the corresponding $TU_{ij}$ and
weight-averaged according to the direction of the galaxy center
($\cos\alpha_{ijk}$). In the above formula, $sgn$ is the sign
function: $sgn(x)$=-1,0,1 for $x<$0, $x$=0 and $x>$0, respectively.

\smallskip\noindent
{\bf $D_{14}$: Monotonicity}.

\noindent
Likewise the previous diagnostic, this one too deals with some
geometrical rule to which the intensities inside $TUs$ should obey. In
particular, in this case we consider the fraction of the total galaxy luminosity
coming from pixels ($i,j$) for which $F_C(i,j)$ has a monotonic behaviour 
in all the four directions of $TU_{ij}$ illustrated in Figure~\ref{pixsqr}. Again, 
the greater the amount of structural features (late--type galaxies), the lower 
the expected fraction. In formula:
$$D_{14}=1+{\displaystyle\sum_{i,j=2}^{N-1}{f_{ij}\times sgn\big{\{}\sum_{k=1}^4{(\delta f_{ij}^{k\pm}-\delta f_{ij}^{k-}-\delta f_{ij}^{k+})\big{\}}}}\over\displaystyle{\sum_{i,j=2}^{N-1}{f_{ij}}}}$$
where:

\noindent
$\delta f_{ij}^{k\pm}$=$\vert f_{ij}^{k-}-f_{ij}^{k+}\vert$, 
$\delta f_{ij}^{k+}$=$\vert f_{ij}^{k+}-f_{ij}\vert$ and 
$\delta f_{ij}^{k-}$=$\vert f_{ij}^{k-}-f_{ij}\vert$.

\medskip\noindent
{\bf $D_{15}$: Alignment}.

\noindent
In defining the next two diagnostics it is convenient to convert the
pixel coordinates in the reference system of the circularized ellipse,
whose ellipticity and position angle ($\varepsilon,\theta$) are those
previously determined (Section~\ref{preproc}) and assumed to be the
global geometrical parameters of the galaxy:

\smallskip
$x_{ij}=[(i-i_c)\cos\theta+(j-j_c)\sin\theta]\times \sqrt{1-\varepsilon}$;

$y_{ij}=[(j-j_c)\cos\theta+(i-i_c)\sin\theta]$/$\sqrt{1-\varepsilon}$;

$z_{ij}=F_C(i,j)$.

\smallskip\noindent
Using this system of coordinates, the new diagnostic ($D_{15}$) tries
to quantify to which degree the local maximum intensity gradient is
aligned with the galaxy center over the whole galaxy body.  Again, the
presence of small scale structures in late--type galaxies should imply
a lower degree of alignment of the local maximum gradient toward the
galaxy center with respect to early--type galaxies.  To quantify such
degree of alignment, the points in the space ($x,y,z$) corresponding
to the nine pixels of each $TU_{ij}$ have been linearly interpolated
($\chi^2$) by the function: $Z_{ij}(x,y)=a_{ij}+b_{ij}x+c_{ij}y$. Then, for
each $TU_{ij}$, we have computed the angle $\phi_{ij}$ between two
planes, both passing through the point
[$x_{ij},y_{ij},Z_{ij}(x_{ij},y_{ij})$] and parallel to the $z$ axis. The
first plane contains the galaxy center (origin of the new coordinates
system), while the second one is parallel to the line of maximum
intensity gradient in the previous linear interpolation. Finally, the
cosines of the angles $\phi_{ij}$ have been weight-averaged over the
whole frame, according to the local intensity. In formula:
$$D_{15}={\sum_{i,j=2}^{N-1}{z_{ij}\times\cos\phi_{ij}}\over{\sum_{i,j=2}^{N-1}{z_{ij}}}}$$
where the cosines:
$$\cos\phi_{ij}={{b_{ij}x_{ij}+c_{ij}y_{ij}}\over{\sqrt{(x_{ij}^2+y_{ij}^2)(b_{ij}^2+c_{ij}^2)}}}$$
turn out to be positive if the function $Z_{ij}(x,y)$ increases toward the galaxy center
at the point ($i,j$), while they are negative in the opposite case.

\medskip\noindent
{\bf $D_{16}$: Intercept Angle}.

\noindent
In defining the previous diagnostic we have introduced the local
planes $Z_{ij}(x,y)$ interpolating the nine points ($x,y,z$) of each
$TU_{ij}$. We noted that in early--type galaxies the line of maximum
gradient of these planes should be (on average) more oriented towards
the center than in the case of late--type galaxies and that such alignment
should be more and more pronounced at increasing the intensity and
decreasing the distance from the galaxy center. Here we note that, in
addition, the average values of the intercept ($a_{ij}$) between these planes and
the $z$ axis should be higher in early--type than in late--type
galaxies, where the irregular intensity surface should make almost
randomly distributed both the orientations and the intercept levels of
the planes, especially in the intermediate and outer regions of
galaxies. In order to quantify such average intercept level avoiding 
computational divergences, we actually prefer to deal with the cosines of the
angles ($\beta_{ij}$) between the $z$ axis and the stright lines
connecting the points ($x_{ij},y_{ij},0$) and ($a_{ij},0,0$). Since
the values of $\cos\beta_{ij}$ obviously depend on the units used to
measure intensities and radii, it is necessary to normalize both
quantities. We decided to normalize the intensities at
$z_c=F_C(i_c,j_c)$ (intensity of the galaxy center) and the radii at:
$R_{thr}=\sqrt{A_{thr}/\pi}$ (equivalent radius of the threshold
area). After some algebra, the new diagnostic, that is the normalized,
average $\cos\beta$ is defined as follows:
$$D_{16}={\sum_{i,j=2}^{N-1}{W_{ij}\times a_{ij}/\sqrt{a_{ij}^2+d_{ij}^2K^2}}\over{\sum_{i,j=2}^{N-1}{W_{ij}}}}$$
where: $d_{ij}=\sqrt{x_{ij}^2+y_{ij}^2}$ is the circularized distance from
the galaxy center, $K=z_c/R_{thr}$ is the normalization factor
and $W_{ij}=z_{ij}/d_{ij}$ are the weighting factors, which increase
at increasing the local intensity and at decreasing the
distance from the galaxy center.

\medskip\noindent
{\bf $D_{17}-D_{20}$: Intensity distribution moments}.

\noindent
The last four diagnostics simply concern the moments of the intensity
distribution within the galaxy frame. In particular, we consider the
median ($D_{17}$), the standard deviation ($D_{18}$), the skewness 
($D_{19}$) and the kurtosis ($D_{20}$) of the distribution of the pixel
intensities, normalized to their average value.

\section{The Maximum Likelihood technique}\label{appml}

   \begin{figure*}
   \centering
   \includegraphics[width=15truecm]{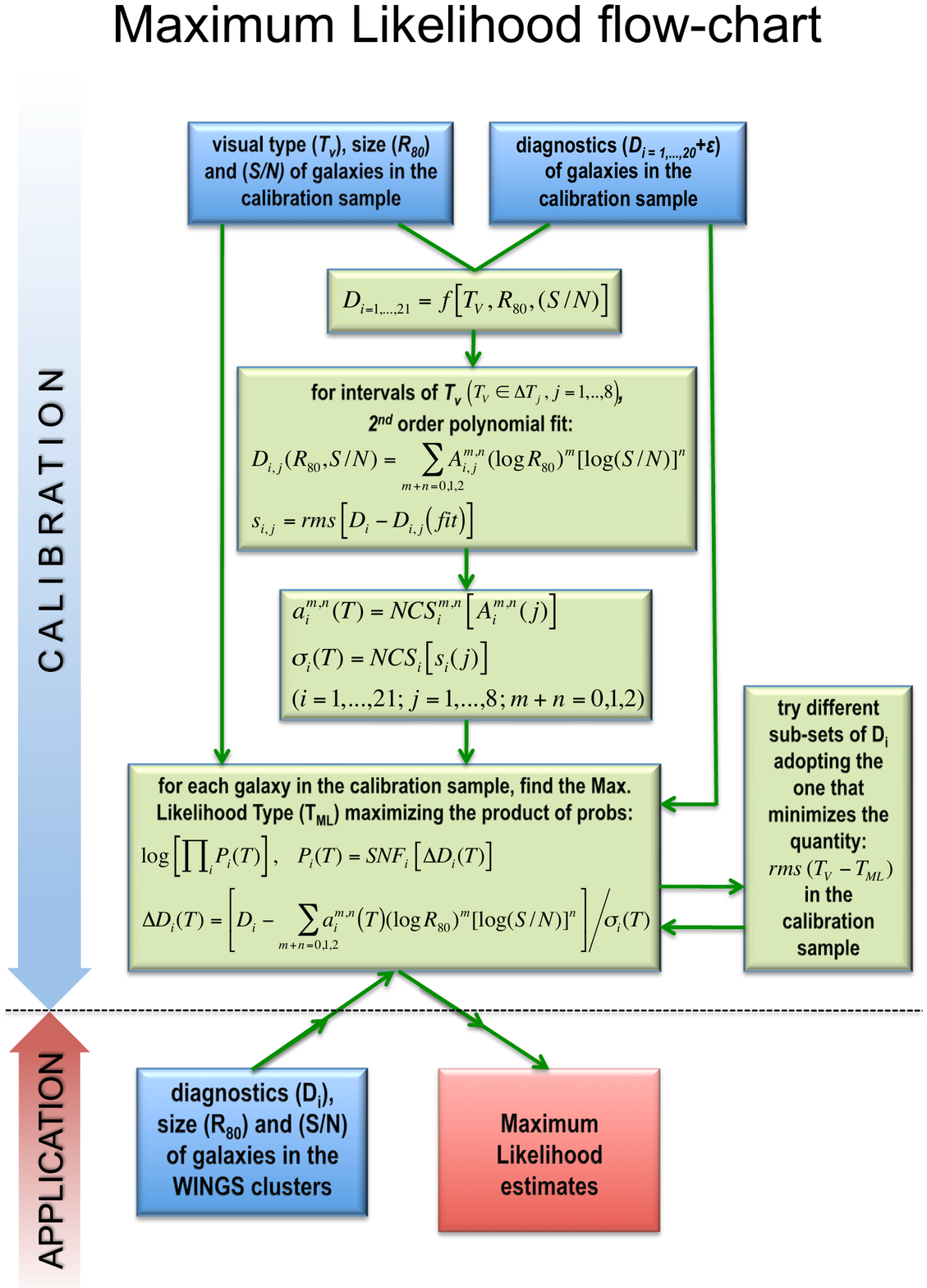}
      \caption{Flow-chart of the MORPHOT Maximum Likelihood tool}
         \label{FC2}
   \end{figure*}

The flow-chart in Figure~\ref{FC2} should help the reader in this
section, leading her/him through the various steps of this
semi-analytical technique. As already mentioned in
Section~\ref{secdep}, besided obviously depending on the visual
morphological type (Figure~\ref{parT}), the diagnostics $D_i$
turn out to depend on both the seeing normalized size
[$\log(R_{80}/FWHM)$] and the signal-to-noise ratio [$\log(S/N)$]
(Figure~\ref{parRSN}). Therefore, prior to use them to
gauge the morphology of galaxies, we have to remove these two
dependences, thus picking out the net dependence of each diagnostic on
the morphological type. 
 
\begin{figure*}
   \vspace{-2.0truecm}
  \includegraphics[width=19truecm]{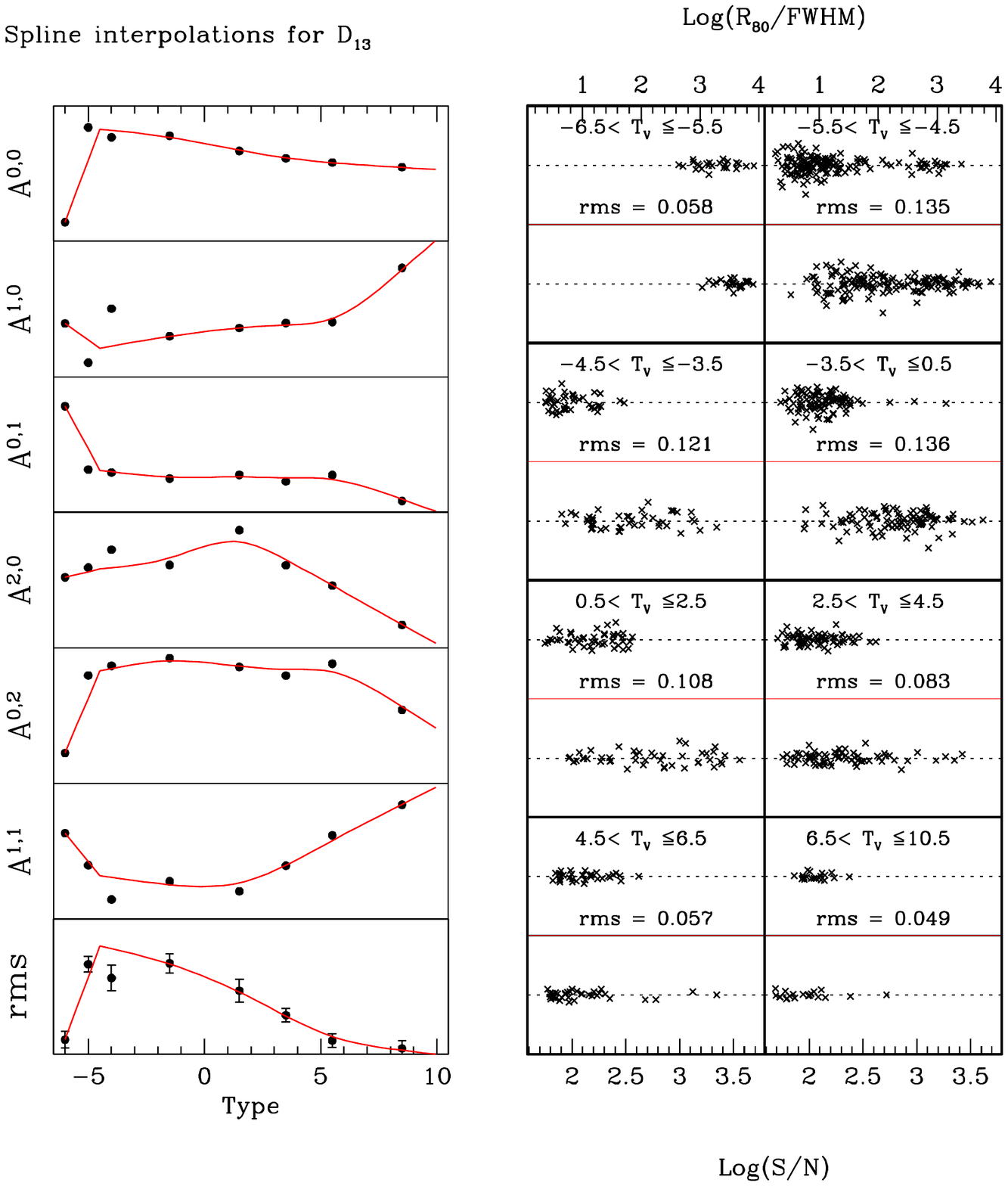}
  \vspace{-5.0truecm}
  \caption{Example of the fitting procedure performed in order to
remove the dependence of morphological diagnostics on the
size ($R_{80}$) and signal to noise ratio ($S/N$). The case
of the diagnostic $D_{13}$ is illustrated in the figure.
{\it Left panels}: Natural Cubic Spline (NCS) fitting of the polynomial
coefficients $A_{13}^{m,n}$ and of the scatter $\sigma_{13}$ as a 
functions of the morphological type (details in the text).
{\it Right panels}: Residuals of the (polynomial+spline) fits illustrated
in the left panels with respect to both $\log(R_{80}/FWHM)$ and $\log(S/N)$.
In particular, the upper panels inside each $\Delta T_j$ tick line box
show the residuals as a function of $\log(R_{80}/FWHM)$ (units at the top
of the figure), while the lower panels in each tick line box show the
corresponding residuals as a function of $\log(S/N)$ (units at the bottom
of the figure).}
   \label{MLdep}
  \end{figure*}

To this aim the whole range of morphological types in the calibration
sample has been divided into eight, nearly omogeneous intervals (see
the plots on the right side of Figure~\ref{MLdep}). For each interval
$\Delta T_j$ (j=1,...,8), the dependences of all the diagnostics 
$D_i$ (i=1,...,21) on $R_{80}$ and $S/N$, have been represented
(through least square fitting) by the following second order
polynomial functions:
$$D_{i,j}(R_{80},S/N)=\hspace{-0.4truecm}\displaystyle\sum_{m+n=0,1,2}{\hspace{-0.4truecm}A_{ij}^{m,n} \log(R_{80}/FWHM)^n \log(S/N)^m}$$
thus producing, for each $D_i$ and for each $\Delta T_j$, the
coefficients $A_{ij}^{m,n}$ and the $rms$ scatters $s_{i,j}$ of the
residuals with respect to the fits. Then, for each $D_i$, these
quantities have been interpolated, as a function of the morphological
type, with Natural Cubic Splines ($NCS$):
$$a_i^{m,n}(T)=NCS_i^{m,n}[A_i^{m,n}(j)];\hspace{0.3truecm}\sigma_i(T)=NCS_i[s_i(j)]$$
An example of this procedure is shown in the plots on the left side of
Figure~\ref{MLdep}, which is referred to the diagnostic
$D_{13}$, defined in Appendix~\ref{appdia}. Because of
the evident discontinuity between cD and E galaxies, the first bin
($\Delta T_1$) has been excluded from the spline fitting. Such
discontinuity is actually found also for the coefficients $A^{m,n}$ of
the other diagnostics and reflects their peculiar behaviour for
$T_M$=$-6$ (see Figure~\ref{parT}). It is likely telling us once again
that cD galaxies are quite apart from Es and, in general, from all
'normal' galaxies, thus suggesting they are the product of a peculiar
evolutionary path \citep{fasa10}. The right side of Figure~\ref{MLdep}
shows (again in the case of $D_{13}$) the residuals of the above
mentioned polynomial fits with respect to both $\log(R_{80}/FWHM)$
(upper panels inside each $\Delta T_j$ heavy box) and $\log(S/N)$
(lower panels in the same boxes).

The previous formulae allow us to remove the dependence of diagnostics
on $R_{80}$ and $S/N$. Consider now a galaxy for which we know the
diagnostics $D_i$ and the quantities $R_{80}/FWHM$ and
$S/N$. For any given morphological type $T$ and for each diagnostic,
the deviations of the actual values of $D_i$ with respect to those
obtained from the above fitting procedure, normalized by the 
corresponding expected scatter, can be expressed as follows:
$$\Delta D_i(T)={{D_i-\hspace{-0.4truecm}\displaystyle\sum_{m+n=0,1,2}\hspace{-0.4truecm}a_i^{m,n}(T) \log(R_{80}/FWHM)^n \log(S/N)^m}\over{\sigma_i(T)}}$$

  \begin{figure}
   \vspace*{-1.5truecm}
   \centering
   \hspace*{-1truecm}
   \includegraphics[width=10truecm]{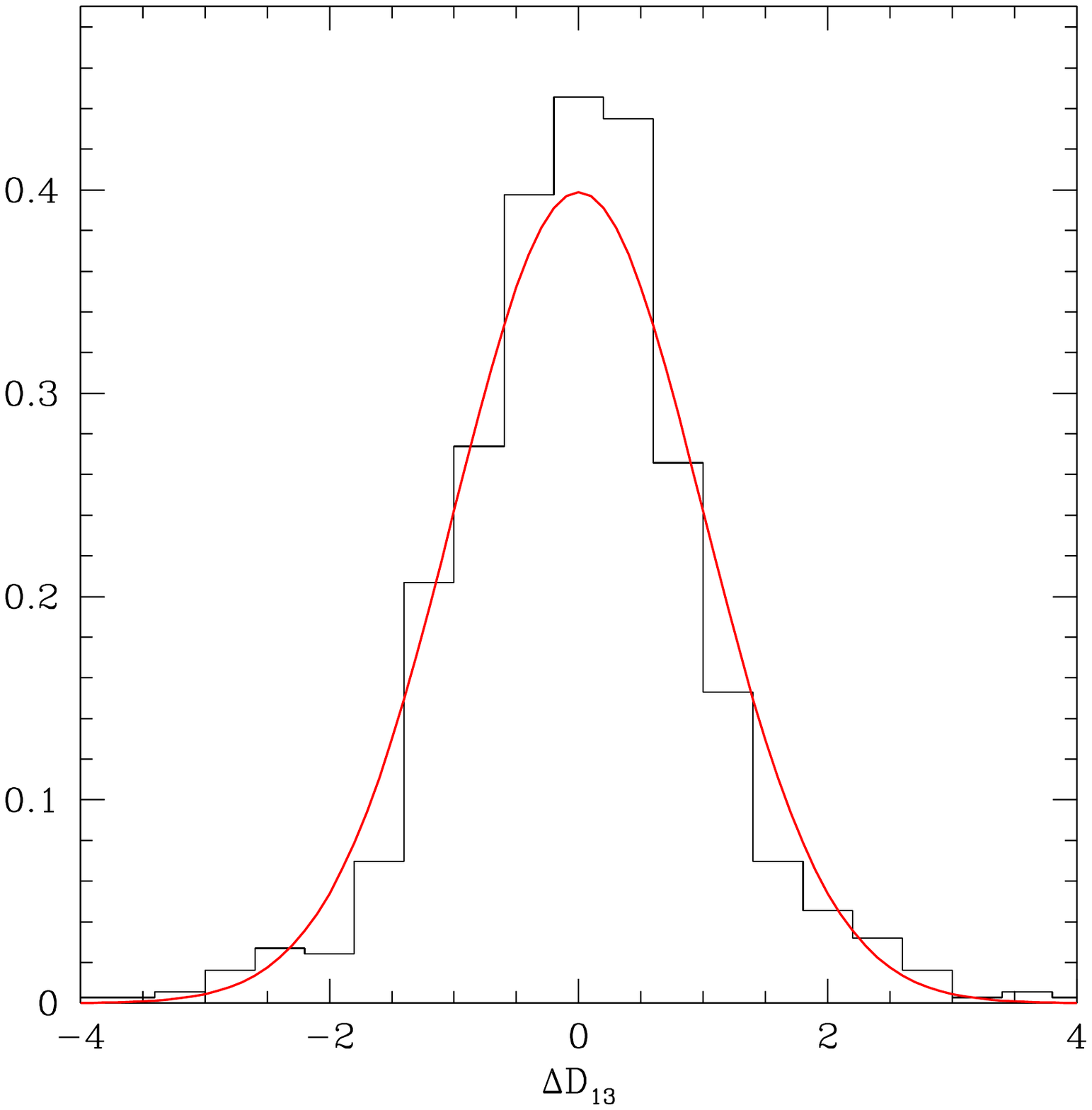}
   \vspace*{-2truecm}
      \caption{Distribution of the deviates of the morphological diagnostics
of the calibration galaxy sample with respect to the polynomial+spline fitting
(details in the text) performed in order to remove their dependence on the
size ($R_{80}$) and signal to noise ratio ($S/N$).}
         \label{MLres}
   \end{figure}

If we assume the distributions of the deviates to be normal (a fair
assumption, in our case; see Figure~\ref{MLres}), the probability that
the actual value of the diagnostic $D_i$ is found for a galaxy of
morphological type $T$, can be expressed by: 
$$P_i(T)=SNF_i[\Delta D_i(T)]$$
($SNF[x]$=$e^{-x^2/2}/\sqrt{2\pi}$ is the Standardized Normal
Function) and the Maximum Likelihood (ML) probability that the
morphological type $T$ is associated to the actual set of diagnostics
$D_i$ ($i$=1,...,21) can be obtained through the product of the
individual probabilities:
$$P(T)=\prod_{i=1,21} P_i(T)$$
We actually prefer to use the logarithmic form:
$$\log[P(T)]=\hspace{-0.2truecm}\displaystyle\sum_{i=1,21}\hspace{-0.2truecm}w_i\times\log[P_i(T)]$$
where $w_i$ are binary weighting factors (0/1) which determine the
inclusion of each diagnostic in the final adopted set of $D_i$ (see
Section~\ref{mltech} and Figure~\ref{dialoop}).

  \begin{figure*}
   \vspace*{-1truecm}
   \hspace*{-1truecm}
   \includegraphics[width=16truecm]{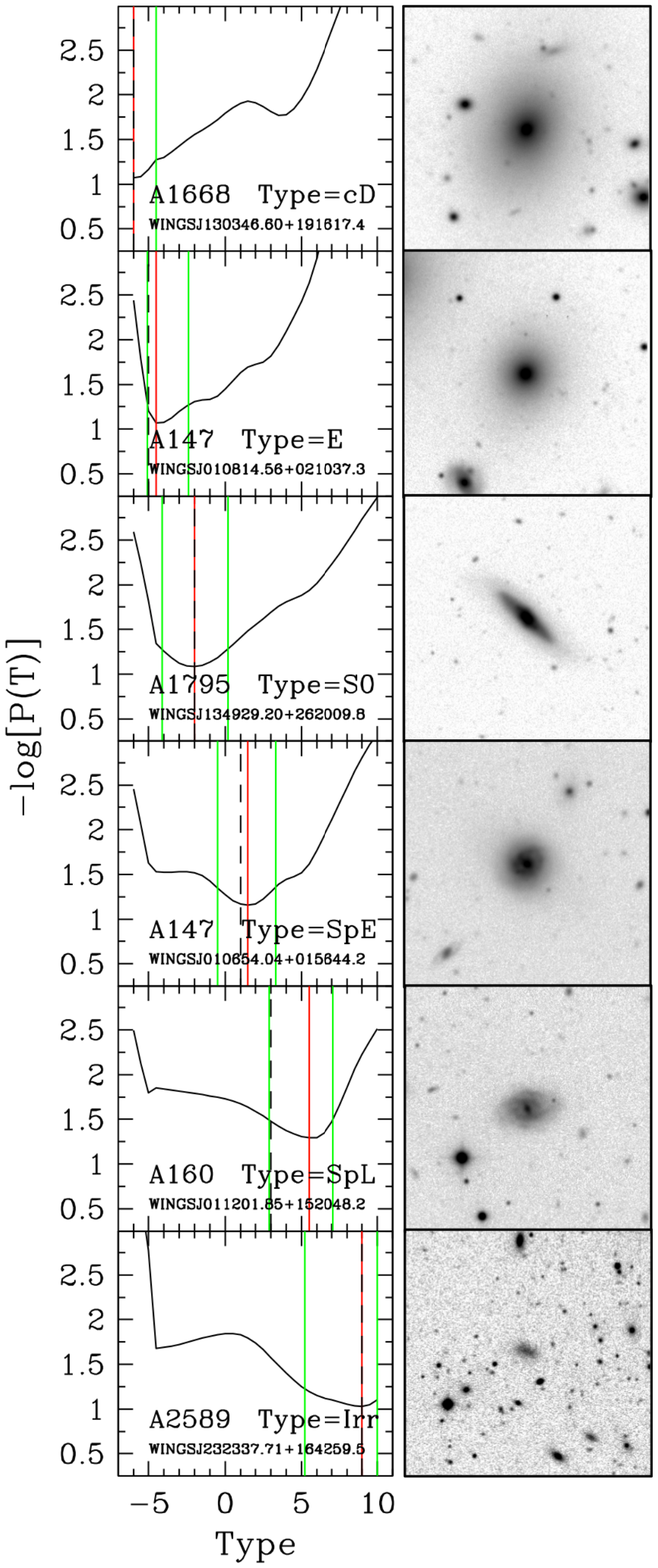}
   \vspace*{-1truecm}
      \caption{Examples of the Maximum Likelihood minimization for
WINGS galaxies of different 'broad' morphological classes.}
         \label{MLexa}
   \end{figure*}

We compute the above probability from $T$=$-6.5$ to $T$=10.5 (step
0.1) and we assume the ML morphological estimator to be the value of
$T$ which minimizes the $-\log[P(T)]$.  Figure~\ref{MLexa} shows some
examples of the behaviour of the ML probability as a function of the
morphological type for typical galaxies belonging to the six above
defined broad morphological classes.  The red lines in each left side
figure mark the minimum values of $-\log[P(T)]$, i.e. the adopted
MORPHOT-ML morphological types $T_{ML}$, while the dashed lines
indicate the corresponding visual estimates $T_V$.  Conventionally, we
should obtain the confidence intervals of our ML estimates by the
change in $T$ necessary to decrease $P(T)$ by $\Delta P$=0.5 from its
value at the maximum (in logarithm: $\Delta\log P\sim$0.3). In
practics, numerical simulations of the MORPHOT-ML technique and visual
inspection of several real cases suggest that a value of $\Delta\log
P\sim$0.2 provides more realistic boundaries to the morphological
estimates. The green lines in Figure~\ref{MLexa} set the confidence
intervals obtained in this way.
Note in the figure the discontinuity between cD and E galaxies 
already shown in Figure~\ref{MLdep} and commented before.
Note also the discrepancy between visual and ML estimates in the
case of the galaxy in Abell~160, which has been visually classified
early-spiral and automatically classified late-spiral ($\Delta T\sim$2.5). 

\section{The Neural Network technique}\label{appnn}

The architecture and strategy adopted to produce a Neural Network
estimator for the morphological classification are very similar to
that described in \citet{vanz04}.  Here be briefly recall the
method.

\subsubsection{Architecture}

We adopt the classical feed-forward multilayer
perceptron Neural Network (MLP, \citealt{bish95}) with four
layers, each one made of 23, 20, 20 and 1 nodes. The first layer
(named input layer) receives 23 input parameters for each galaxy
(i.e. the diagnostics $D_i$, $i$=1,...,20 as described in
Appendix~\ref{appdia} and the global quantities $\varepsilon$,
$\log[R_{80}/FWHM]$ and $\log[S/N]$), while the single output node
(output layer) produces the morphological evaluation. The other two
layers of 20 nodes each are called hidden layers. Each node of a layer
is connected to all the nodes of the next layer.

\subsubsection{Training}

We use the supervised learning method, i.e. the NN is
trained with examples. The training set is composed by the 926
galaxies in the calibration sample, each one visually classified
($T_V$).  Each example presented to the NN is a pair of arrays, one
contains the set of diagnostics (described in Appendix~\ref{appdia})
with 23 components and the second array is the {\it targeted}
morphology (single component), $T_V$.  The learning
algorithm\footnote{Here we have used the {\it back-propagation}
  algorithm with its generalized {\it delta rule} version (see
  \citealt{vanz04}.)} modifies the strength of the connections between
nodes (called weights) in order to force the association between the
input and the correct output. It translates to a minimization of a
merit function as a function of the set of weights.

Once a suitable set of weights is determined, it is frozen and stored
for evaluation of new (never seen) galaxies. There are various
techniques to identify the best set of weights, i.e. those that offer
the so called best {\it generalization power} and that avoid the
over-fitting problem.
Here we adopt the method based on a committee of neural networks (30
MLPs have been used) as described in \citet{vanz04}.  This
method reduce the variance in the predictions maintaining a relatively
small bias \citep{bish95}.  Each member of the committee has been
trained with different initial conditions (e.g. initial random
distribution of weights, random sequence of examples presented to the
NN, bootstrapping of examples), that produce different histories of
training and 30 different set of weights.  Therefore, for each galaxy,
the committee produces 30 estimations of the morphology ($T_{NN}^{i}$,
$i$=1-30). From this distribution we extract the median(mean)
$T_{NN}$ of the $T_{NN}^{i}$ and the central 68\% interval (16 and 84
percentiles). With this method we can associate a statistical
uncertainty to the predicted value $T_{NN}$.

It is clear that the outlined weighting of the connections, inherent to
this methodology, automatically provides the selection of the
most significant diagnostics, thus making useless the preliminar
(empirical) choice we performed in the case of the ML technique.

\end{document}